\documentclass[12pt]{article}
\usepackage[dvips]{graphicx}
\usepackage{amssymb}
\usepackage{color}
\definecolor{Blue}{rgb}{0.3,0.3,0.9}
\definecolor{Red}{rgb}{1,0,0}
\definecolor{Green}{rgb}{0,1,0}
\newcommand{\be}{\begin{equation}}
\newcommand{\ee}{\end{equation}}
\newcommand{\bea}{\begin{eqnarray}}
\newcommand{\eea}{\end{eqnarray}}

\begin{document}
\begin{titlepage}
\begin{flushright}
DCPT-05/41
\end{flushright}
\smallskip

\begin{center}
{\bf \large Exact Results for Evaporating Black Holes in Curvature-Squared
Lovelock Gravity\,:\\[1.5mm] Gauss-Bonnet Greybody Factors}
\vskip 1.5 cm
{\bf J. Grain,}$^{1}$ {\bf A. Barrau}$^{1}$ and {\bf P. Kanti}$^{2}$
\vskip 0.3cm
$^{1}$ Laboratory for Subatomic Physics and Cosmology,\\
 Joseph Fourier University, CNRS-IN2P3,\\
 53, avenue des Martyres, 38026 Grenoble cedex, France
 
 \bigskip
 $^{2}$ Department of Mathematical Sciences, University of Durham,\\
 Science Site, South Road, Durham DH1 3LE, United Kingdom
\vskip 0.3cm

\end{center}
\date{\today}

\vskip 1cm
\small
\begin{center}
{\bf Abstract}
\end{center}

\begin{quote}
Lovelock gravity is an important extension of General Relativity that provides a
promising framework to study curvature corrections to the Einstein action, while
avoiding ghosts and keeping second order field equations. This paper derives
the greybody factors for D-dimensional black holes arising in a theory with a
Gauss-Bonnet curvature-squared term. These factors describe the non-trivial
coupling between black holes and quantum fields during the evaporation process:
they can be used both from a theoretical viewpoint to investigate the intricate
spacetime structure around such a black hole, and for phenomenological purposes
in the framework of braneworld models with a low Planck scale. We derive exact
spectra for the emission of scalar, fermion and gauge fields emitted on the
brane, and for scalar fields emitted in the bulk, and demonstrate how the
Gauss-Bonnet term can change the bulk-to-brane emission rates ratio in favour
of the bulk channel in particular frequency regimes.
\end{quote} 
\normalsize
\date{\today}

\end{titlepage}

\section{Introduction}

In any attempt to perturbatively quantize gravity as a field theory, higher-derivative
interactions must be included in the action. Such terms also arise in the effective
low-energy action of string theories. Furthermore, it has been argued
that higher-derivative gravity theories are intrinsically attractive as in many cases
they display features of renormalizability and asymptotic freedom. Among such approaches,
Lovelock gravity \cite{love} is especially interesting as the resulting equations of
motion contain no more than second derivatives of the metric, and as they are free
of ghosts when expanding around flat space. Among the dimensionally extended Euler
densities used in the Lovelock Lagrangian, it is reasonable to consider that the
four-derivative Gauss-Bonnet term is the dominant correction to the Einstein-Hilbert
action \cite{zie}. The action therefore reads as\,:
$$S_{GB}=\frac{1}{16\pi G}\int d^{D}x\sqrt{-g}\left[
R+ \alpha (  R_{\mu\nu\rho\sigma} R^{\mu\nu\rho\sigma} -
4 R_{\mu\nu} R^{\mu\nu} + R^2 )\right]\,,$$
where $\alpha$ is a coupling constant of dimension (length)$^2$, and $G$ the
$D-$dimensional Newton's constant defined as $G=1/M_*^{D-2}$ in terms of the
fundamental Planck scale $M_*$. 
In four dimensions, the Gauss-Bonnet term is a total derivative, but it becomes dynamical
for $D>4$. The thermodynamical properties of such black holes have been already studied 
both in asymptotically flat \cite{myers} and curved \cite{cai} spacetimes. Unfortunately,
the temperature and entropy which have been derived are not sufficient to describe the
detailed evaporation spectrum through the emission of Hawking radiation\,: the exact 
coupling between the black hole and the quantum fields, {\it i.e.} the greybody factors, 
must be accurately computed. To date, the greybody factors have been obtained for
D-dimensional Schwarzschild \cite{kanti2,kanti3,harris} (for a detailed review,
see \cite{kanti1}),  Reissner-Nordstrom \cite{Jung1}, Kerr \cite{ida,harris2,Jung2,duffy} 
and Schwarzschild-de-Sitter \cite{kgb} black holes, for various types of emitted fields.

Furthermore, it has been recently pointed out that small black holes could be
formed at next-generation colliders during trans-planckian particle collisions
\cite{creation} if the Planck scale is of order a TeV, as is the case in some
scenarios postulating the existence of extra dimensions \cite{ADD}. This idea
has driven a considerable amount of interest in the framework of collider physics
\cite{colliders}, as well as in cosmic-ray physics \cite{cosmic} as the same
phenomenon could also occur during ultrahigh-energy neutrino interactions
in the atmosphere of the earth. Most works consider those black holes as
described by higher-dimensional generalizations of 4-dimensional line-elements
that follow in the context of Einstein's linear theory of gravity. Studying the
experimental consequences of the Gauss-Bonnet term, when included in the D-dimensional
action, is a very challenging path strongly connected to experimentally probing quantum 
gravity. With arguments based solely on the dynamics of the temperature, it has been
pointed out that for realistic values of the Gauss-Bonnet coupling constant and a low
enough Planck scale, the LHC should be able to discriminate between a pure Einstein
action and a second order Lovelock action \cite{bga} upon the successful detection of
the Hawking radiation emission spectrum from such a black hole. 

To confirm this stimulating hypothesis it is nevertheless necessary to calculate
the exact Gauss-Bonnet greybody factors values that have not been computed to date.
The aforementioned work \cite{bga} used an approximate expression for the greybody
factors to compute the emission spectrum from a Schwarzschild-Gauss-Bonnet black
hole; this approximate expression ignored altogether its dependence on the energy
and spin of the emitted particle as well as its dependence on the Gauss-Bonnet
coupling constant itself. In this work, we calculate through numerical
analysis the exact values of the Gauss-Bonnet greybody factors and the corresponding
emission spectra for scalar, fermion and gauge boson fields restricted to live
on our 4-dimensional brane. The spectrum for the emission of scalar fields in the
bulk will also be studied, and the bulk-to-brane ratio will be accurately calculated.
The dependence of the spectra on the Gauss-Bonnet coupling constant will be studied
in detail, and footprints of its value would be looked for in both bulk and brane
emission channels. As we will see, the implementation of the exact Gauss-Bonnet
greybody factors will lead to modifications in the Hawking radiation emission
spectra compared to the ones derived in the pure Schwarzschild case, and can
also change the so-far derived results in the literature for the bulk-to-brane
emission rates ratio.

In what follows, we will work in the context of the scenario with Large Extra
Dimensions \cite{ADD}, and assume that the higher-dimensional spacetime is empty
and thus flat. Also, the horizon value $r_H$ of the produced black holes will be
assumed to be much smaller than the (common) size ${\cal R}$ of the extra dimensions;
under this assumption, the extra spacelike dimensions can be considered to be
non-compact as the usual 3-dimensional ones. The energy of the emitted particle 
$\omega$ will be taken to be always much smaller than the black hole mass $M_{BH}$
so that no significant back-reaction to the black hole background will take place
after the emission of a particle. The frequency of the emitted particle must also
be bounded from below, {\it{i.e.}} $\omega \gg 1/{\cal R}$; this guarantees that the
wavelength of the particle is much smaller than the size of the extra dimensions
and the particle can therefore be considered as a genuinely higher-dimensional one.
Although the upper limit on the frequency of the black hole will be always
ensured in our analysis, for the sake of presenting complete spectra, the
lower bound will be ignored and restored at the end of our analysis. 

Section 2 of this paper sets the general framework for our analysis. The absorption
cross-sections, {\it{i.e.}} the greybody factors, are then numerically computed for
particles on the brane and in the bulk in sections 3 and 4, respectively. 
Section 5 presents the final Hawking radiation emission spectra for both
brane and bulk channels, and the bulk-to-brane emission rates ratios is 
explicitly given. We discuss our results and summarise our conclusions in
section 6. The numerical values of the greybody factors produced in this
analysis, that might be useful for experimental searches, can be downloaded
from the URL given in \cite{web}.


\section{General Framework}
A higher-dimensional, neutral, static, spherically-symmetric black hole within 
the framework of a curvature-squared, Gauss-Bonnet gravity is described by the
following line element, first derived in \cite{deser} and then extended to a
different topology in \cite{cai}:
\begin{equation}
	ds^2=h(r)\,dt^2-\frac{dr^2}{h(r)}-r^2\,d\Omega^2_{D-2}\,,
	\label{metric-bulk}
\end{equation}
where $d\Omega^2_{D-2}$ is the line-element over a unitary $(D-2)$-dimensional 
sphere and
\begin{equation}
	h(r)=k+\frac{r^2}{2\alpha(D-3)(D-4)}\left(1\pm\sqrt{1+
	\frac{64\pi\alpha(D-3)(D-4)G M_{BH}}{(D-2)\Omega_{D-2}\,r^{D-1}}}\right)
\label{h}
\end{equation}
is the metric function. In the above, $\Omega_{D-2}$ corresponds to the total 
$(D-2)$-dimensional solid angle. 
In \cite{deser}, it was shown that a Schwarzschild-like space-time within 
Gauss-Bonnet gravity exhibits two branches corresponding to the $\pm$ signs 
entering the metric function. The $(+)$-sign branch describes a Schwarzschild
black hole with a negative mass parameter embedded in an Anti-de Sitter Universe 
whereas the $(-)$-sign branch corresponds to an asymptotically flat 
Schwarzschild black hole with a positive mass parameter, nevertheless both
branches are characterized by a positive ADM energy, as it was shown in
\cite{DT}. The authors of Ref. \cite{cai} extended 
the solution of \cite{deser} to different topologies for the black hole's event 
horizon~: for $k=1$, the event horizon is elliptic whereas it 
becomes flat for $k=0$ and hyperbolic for $k=-1$ \cite{birmingham}. 
In this work, we restrict our analysis to the $(-)$-sign branch with an 
elliptic topology for the event horizon, as it corresponds to the usual 
Schwarzschild black hole solution. The $(+)$ branch will be hereafter ignored
and the $k$ parameter will be set equal to +1.

As in higher-dimensional general relativity, Schwarzschild-Gauss-Bonnet 
black holes present one event horizon located at $r_H$ and defined as the 
solution of the algebraic equation $h(r)=0$. The thermodynamical properties of 
those black holes have been extensively studied in \cite{cai}. For non-rotating, 
neutral black holes, the thermodynamics is described by three quantities, all 
parametrized by the horizon radius: the mass $M_{BH}$
\begin{equation}
   M_{BH}=\frac{(D-2)\,\pi^{\frac{D-1}{2}}r_H^{D-3}}
   {8\pi G\,\Gamma\left(\frac{D-1}{2}\right)}\left(1+\frac{\alpha(D-3)(D-4)}{r_H^2}\right)\,,
\label{Mass}
\end{equation}
where $\Gamma$ stands for the Euler Gamma function, the temperature $T_{BH}$ given by 
the surface gravity evaluated at the event horizon radius,
\begin{equation}
	T_{BH}=\frac{(D-3)\,\left[\,r_H^2+ \alpha (D-5)(D-4)\,\right]}
	{4\pi{r}_H\left[\,r_H^2+2\alpha(D-4)(D-3)\,\right]}\,,
\label{temp}
\end{equation}
and the entropy $S_{BH}$, defined by
\begin{equation}
	S_{BH}=\displaystyle\int\frac{dM_{BH}}{T_{BH}}.
\end{equation}

The Hawking radiation emission process by the black hole given in Eq. (\ref{metric-bulk})
takes place in the higher-dimensional space-time with a spectrum given by the
generalization of the four-dimensional one obtained in \cite{hawking}~:
\begin{equation}
\frac{dN_s(\omega)}{dt}=\displaystyle\sum_j\frac{\sigma_{j,s}(\omega)}{\exp{\left({\omega}/{T_{BH}}\right)}-\left(-1\right)^{2s}}\frac{d^{D-1}k}{\left(2\pi\right)^{D-1}}.
\end{equation}
In the above equation, $j$ stands for the total angular momentum quantum number 
and $s$ for the spin of the particle. In this work, we will focus on massless particles
for which $\left|k\right|=\omega$; in this case, the expression for the flux is greatly 
simplified as the phase space term reduces to an integral over the energy of the
emitted particle
\begin{equation}
\frac{d^2N_s(\omega)}{dt\,d\omega}=\displaystyle\sum_j\frac{\sigma_{j,s}(\omega)\,\omega^{D-2}}
{\exp{\left({\omega}/{T_{BH}}\right)}-\left(-1\right)^{2s}}\frac{\Omega_{D-2}}
{\left(2\pi\right)^{D-1}}.
	\label{rad-spec}
\end{equation}
The function $\sigma_{j,s}(\omega)$, usually called ``greybody factor'' or alternatively 
``absorption/emission cross section'', when accurately taken into account, makes the
resulting radiation spectrum more complex than the usual blackbody law.
This additional factor arises because of the non-vanishing probability for the emitted
particles to be reflected on the gravitational potential barrier and re-absorbed by the
black hole. The cross section is computed within a full quantum mechanical framework to
determine first the tunnelling transmission probability $\left|\mathcal{A}_{j,s}\right|^2$
in the aforementioned Schwarzschild-Gauss-Bonnet background. The cross section for a
given angular momentum quantum number is then deduced from the generalization of the
optical theorem \cite{gubser}~:
\begin{equation}
\sigma_{j,s}(\omega)=\frac{2^{D-4}\pi^{(D-3)/2}\Gamma\left((D-3)/2\right)}
{\left(D-4\right)!\,\omega^{D-2}}\times\frac{\left(2j+D-3\right)\left(j+D-4\right)!}
{j!}\left|\mathcal{A}_{j,s}\right|^2\,.
	\label{abs-cros}
\end{equation}

As it is clear from the above, the Hawking radiation emission spectrum is defined
by two quantities, the black hole temperature and the particle's greybody factor.
From Eq. (\ref{temp}), we may see that the temperature has an explicit dependence
on both the Gauss-Bonnet coupling constant and the dimensionality of spacetime.
As we will see, the greybody factor depends, too, upon the underlying gravitational
theory and thus encodes valuable information on both of these parameters; as expected,
it also depends on properties of the emitted particle such as its energy and spin. 
This elaborate dependence of the greybody factor on a number of fundamental parameters
of the theory is the main motivation for performing the exact calculation of its
value and of the corresponding black hole radiation spectrum within the framework
of a $D$-dimensional Gauss-Bonnet gravity.


\section{Radiation on the brane}

\subsection{General equation for particles on the brane}
	Particles confined on the brane propagate into a background whose geometry is induced by the bulk curvature. This induced geometry on the $4-$dimensional brane is simply given by fixing the values of the extra azimuthal angular coordinates, $\theta_i=\pi/2$, and leads to the projection of the $D-$dimensional metric on the $4-$dimensional slice that describes our
world. The projected metric has the form:
\begin{equation}
	ds^2=h(r)dt^2-\frac{dr^2}{h(r)}-r^2\left(d\theta^2+\sin^2{\theta}d\varphi^2\right)\,,
	\label{metric-brane}
\end{equation}
where the metric function $h(r)$ remains the same as in the higher-dimensional line-element
given in Eq. (\ref{metric-bulk}). Consequently, the profile of the curvature along the three non-compact spatial dimensions contains a footprint of the bulk curvature.

The equations of motion describing the behavior of particles with spin restricted to our
$4-$dimensional brane are deduced by using the Newman-Penrose formalism \cite{newman1,newman2}.
Under the assumption of minimal coupling to gravity, these equations can be combined
into one {\it master} equation for particles with spin $s=0,~1/2,~1$. In \cite{kanti1}
(see also \cite{ida}), this master equation was derived for particles confined on a brane embedded in a $D-$dimensional Kerr black hole background. In the limit of zero black hole
angular momentum, the Kerr metric reduces to the Schwarzschild one, and the induced
background on the brane assumes the form given in Eq. (\ref{metric-brane}). Using the
following factorized ansatz for the propagating field:
\begin{equation}
\Psi_s(t,r,\theta,\varphi)=e^{-i\omega{t}}e^{-im\varphi}\Delta^{-s}P_s(r)Y^m_{s,j}(\theta)
	\label{ansatz}
\end{equation}
where $\Delta=hr^2$ and $j$ is the total angular momentum quantum number, the radial part
of the master equation of motion is found to have the form: 
\begin{equation}
\Delta^{s}\frac{d}{dr}\left(\Delta^{1-s}\,\frac{dP_s}{dr}\right)+\left(\frac{\omega^2r^2}{h}
+2is\omega{r}-\frac{is\omega{r}^2h^\prime}{h}-\tilde{\lambda}\right)P_s=0
	\label{master-eq}
\end{equation}
In the above, $\tilde{\lambda}=j(j+1)-s(s-1)$ is derived from the master angular equation satisfied by the spin-weighted spherical harmonics $Y^m_{s,j}$ \cite{y_lm}. 

The aforementioned master equation has been successfully used to derive the Hawking
radiation emission spectrum on the brane for scalar, fermion and gauge boson fields
by a $D-$dimensional Schwarzschild black hole \cite{kanti2,kanti3,harris}, and for
scalar fields by a $D-$dimensional Schwarzschild-de Sitter black hole \cite{kgb}. 
Recently, its complete form, that includes a non-vanishing black hole angular momentum,
was used to study the emission of scalar fields on the brane by a $D-$dimensional 
Kerr black hole \cite{harris2, duffy}. In our case, since the line-element of a
$D-$dimensional Schwarzschild-Gauss-Bonnet black hole has a Schwarzschild-like form,
Eq. (\ref{master-eq}) can be consistently used to study the emission of 
scalar\,\footnote{The equation of motion for a scalar field propagating in a
$D-$dimensional Schwarzschild-Gauss-Bonnet black hole background was recently
studied in \cite{Abdalla} to obtain the evolution of the scalar field in time and
its quasinormal modes.}, fermion and gauge boson fields emitted by such a black
hole on the brane.

\begin{figure}[t]
	\begin{center}
		\includegraphics[scale=0.7]{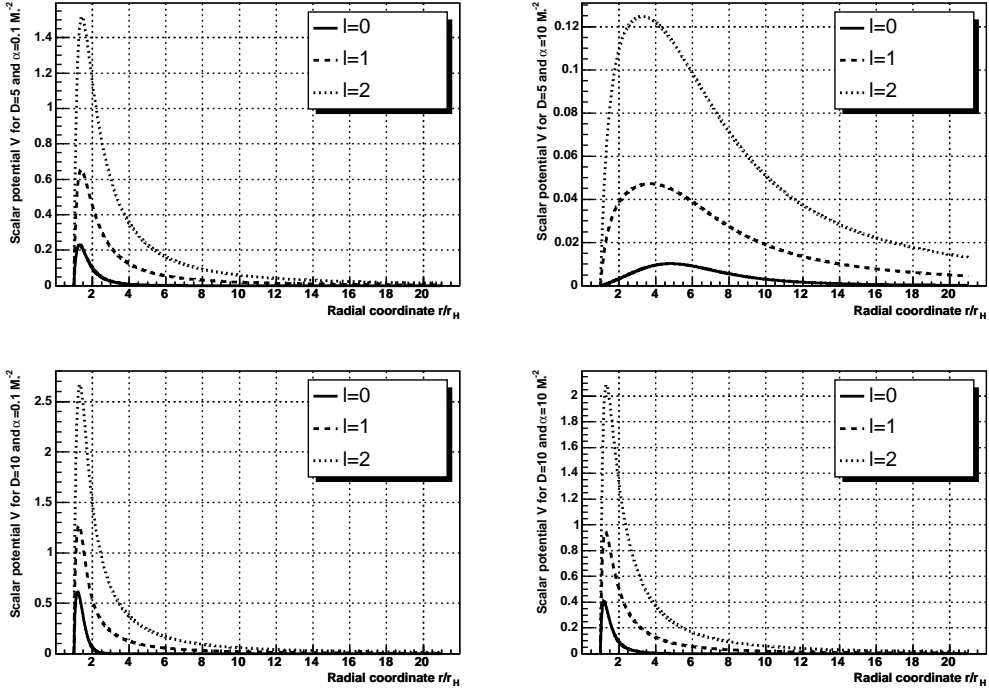}
		\caption{Gravitational potential seen by scalar particles propagating in 
		the background of an induced Schwarzschild-Gauss-Bonnet black hole. 
		{\it Upper Panels:} The number of dimensions is fixed at 5 and the 
		Gauss-bonnet coupling constant at $0.1~M^{-2}_{*}$ on the left and 
		at $10~M^{-2}_{*}$ on the right. {\it Lower Panels:} The number of 
		dimensions is fixed at 10 and the Gauss-bonnet coupling constant at
		$0.1~M^{-2}_{*}$ on the left and at $10~M^{-2}_{*}$ on the right.}
		\label{pot}
	\end{center}
\end{figure}

Writing the master equation (\ref{master-eq}) for scalar particles in an alternative
form allows to validate our approach to be followed in the next section. By using the
tortoise coordinate $r^\star$, defined by $h(r)dr^\star=dr$, and redefining the unknown
radial function by $U(r)=rP_0(r)$, the scalar equation takes a Shr\"odinger-like form:
\begin{equation}
	-\frac{d^2U}{{dr^\star}^2}+h(r)\left(\frac{\ell(\ell+1)}{r^2}
        +\frac{h^\prime(r)}{r}\right)U=\omega^2U\,,
	\label{schro-eq}
\end{equation}
where $\ell$ is the orbital angular momentum. From the above equation, the gravitational
potential seen by scalar particles can be easily read, and it is depicted in Fig. \ref{pot}
for various values of the fundamental parameters of the theory.
This potential, for Schwarzschild-Gauss-Bonnet black holes,
tends to zero both at the black hole's event horizon and at spatial infinity, due to
the vanishing of the metric function in the first case and the asymptotic flatness
of space-time in the second. The corresponding near-horizon and far-field asymptotic
states can be therefore defined in terms of plane waves, a result that allows the
use of the optical theorem for the computation of the absorption cross-section.
Figure \ref{pot} clearly shows that the height of the potential barrier is increasing
for modes with increasingly larger angular momentum number, and a similar effect
arises as the dimensionality of spacetime increases, too. On the other hand, the
height of the barrier is considerably lowered as the value of the Gauss-Bonnet
coupling increases, with the effect being more significant for low values of $D$. 

\subsection{Numerical results for the absorption cross-section}

Equation (\ref{master-eq}) has been attacked analytically in the literature, and
results have been produced for the absorption cross-section (see for instance
Refs. \cite{kanti2, kanti3, kgb}) under the assumption of low-energy emitted particles. 
As the gravitational background, however, becomes more complicated, our ability to
solve this equation analytically is greatly reduced. Moreover, the need to produce
exact results, free from any approximations and valid throughout the energy regime,
dictates the use of numerical analysis. The intricate shape of the metric function
(\ref{h}) in this particular case makes Eq. (\ref{master-eq}) impossible to solve
analytically. Therefore, in this work, the calculation of the absorption cross section
is performed numerically, for particles with spin $s=0,~1/2$ and 1. 

The numerical integration of Eq. (\ref{master-eq}) starts from the black hole horizon,
where appropriate boundary conditions are applied, and extends to spatial infinity. 
In order to deal with finite numbers and avoid divergences at the horizon, the
numerical integration starts at $r_{ini}=r_H+\varepsilon$ instead of $r_H$, where
$\varepsilon$ is small enough to induce a negligible error. As mentioned before,
the asymptotic solution near the horizon is of a plane-wave form. In order to
simplify the numerical analysis, the $P_s$ function and its derivative are fixed,
at $r_{ini}$, to the following values~:
\begin{eqnarray}
		&& P_s(r_{ini})=1  \label{bound1}\\[2mm]
		&&\frac{dP_s}{dr}=-\frac{i\omega}{h(r_{ini})}\,,
\label{bound2}
\end{eqnarray}
which ensure that the no-outgoing modes boundary condition is satisfied, and fix the
flux at the  black hole event horizon to $\left|P_s\right|^2=1$.

The solution of the master equation is then propagated from the event horizon
until a high enough value of the radial coordinate $r$, to be considered as spatial 
infinity. At this asymptotic regime, the wave function must be described by spherical
waves \cite{kanti2}\,:
\begin{equation}
   P_s(r)\simeq{B}_1\,\frac{e^{-i\omega{r}}}{r^{1-2s}}+{B}_2\,\frac{e^{i\omega{r}}}{r}\,,
   \label{asymp-sol}
\end{equation}
where $B_1$ is the amplitude of the ingoing modes and $B_2$ the amplitude of the 
outgoing modes. The numerical results are then compared to the above asymptotic
solution, and a fit is performed in order to extract the amplitudes $B_1$ and $B_2$.
As it is obvious from the above asymptotic solution, for scalar particles with $s=0$,
the solution of the master equation at infinity contains both ingoing and outgoing modes.
For particles with non-vanishing spin though, the complete solution follows by solving 
a set of two equations~: one for the upper component $s=+\left|s\right|$, which 
is mainly composed of ingoing modes, and one for the lower component $s=-\left|s\right|$,
which is mainly composed of outgoing modes \cite{cvetic}. The absorption probability
can be found through the ratio of the incoming flux at infinity over the one at the
horizon, therefore only the solution for the ingoing modes is necessary for our study,
and thus only the upper component of Eq. (\ref{master-eq}) is solved.  The absorption
cross-section is then easily computed in terms of the absorption probability, through
Eq. (\ref{abs-cros}) evaluated for $D=4$ to account for brane emission~:
\begin{equation}
\sigma_j(\omega)=\frac{\left(2j+1\right)\pi}{\omega^2}\left|\mathcal{A}_j\right|^2.
\end{equation}
To obtain the final absorption cross-section, all the angular momentum contributions
have to be summed. Fortunately, as it is shown in Fig. \ref{pot}, the gravitational
potential barrier increases with $j$, and the high angular momentum contributions 
are strongly suppressed. The sum is therefore performed until rank $j=m$ such 
that the error due to higher momentum contributions becomes irrelevant.

In the case of particles with non-vanishing spin, another difficulty arises due to
the very small value of the outgoing modes in the upper component of the wave function.
As these modes are extremely small at the event horizon, a tiny error in the numerical
solution can easily be  mixed with an outgoing solution and can consequently contaminate
the ingoing one. As a result, the numerical integration is not stable as it strongly
depends on the boundary conditions. This contamination does not arise for fermions,
but it becomes significant for gauge boson fields, and the absorption cross-section 
cannot be robustly computed by integrating Eq. (\ref{master-eq}). A solution to 
overcome this problem has been proposed in \cite{harris} and consists in solving
the equation of motion for a new unknown radial wave function $F(y)$, defined as 
$P_1=yF(y)e^{-i\omega{r}^\star}$, with $y=r/r_H$. In terms of these new variables, the 
wave equation for gauge boson fields becomes~:
\begin{equation}
	hy^2\frac{d^2F}{dy^2}+2y(h-i\omega{r}_Hy)\frac{dF}{dy}-j(j+1)F=0
\end{equation}
with the following boundary conditions having replaced the ones in Eqs.
(\ref{bound1})-(\ref{bound2})\,: $F(1)=1$ and $dF/dy(1)=ij(j+1)/2\omega{r}_H$.

\subsubsection{Absorption cross-section for scalars}
	
The absorption probability is defined as the ratio of the ingoing flux ${\cal F}$ at
the event horizon over the one at infinity. For scalar particles, the flux is given
by the radial part of the conserved current $J^\mu=2hr^2 {\mathcal Im}
{\bigl(\Psi_0\partial^\mu\Psi^\dag_0\bigr)}$, which leads to 
$\mathcal{F}=2hr^2\mathcal Im{\Bigl(\frac{P_0}{r^2}\frac{d}{dr}r^2P^\dag_0\Bigr)}$. 
The absorption probability is found to have the form
\begin{equation}
\left|\mathcal{A}_j\right|^2=r_H^2\,\left|\frac{1}{B_1}\right|^2=
                            1-\left|\frac{B_2}{B_1}\right|^2\,,
	\label{scal-coef}
\end{equation}
where, in the middle part, we have used the fact that the ongoing flux at the
black hole horizon has been fixed to 1 by the boundary condition (\ref{bound1}).
The last part of the above equation offers an alternative but equivalent way of
calculating $\left|\mathcal{A}_j\right|^2$ through the reflection probability
given in terms of the amplitudes of the ingoing and outgoing modes at infinity.

\begin{figure}
	\begin{center}
		\begin{tabular}{cc}
			\includegraphics[scale=0.35]{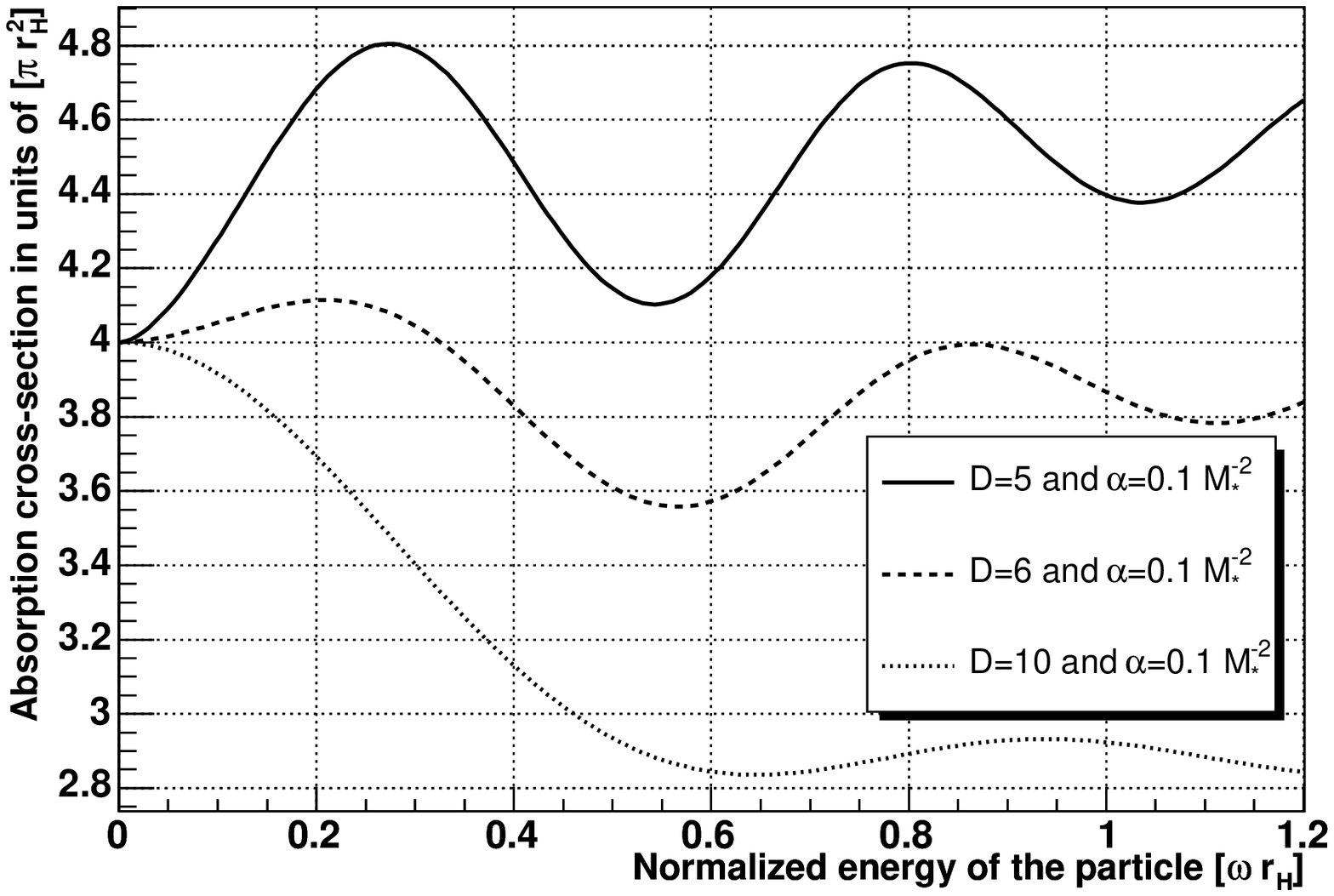}&
 \hspace*{-0.5cm}\includegraphics[scale=0.35]{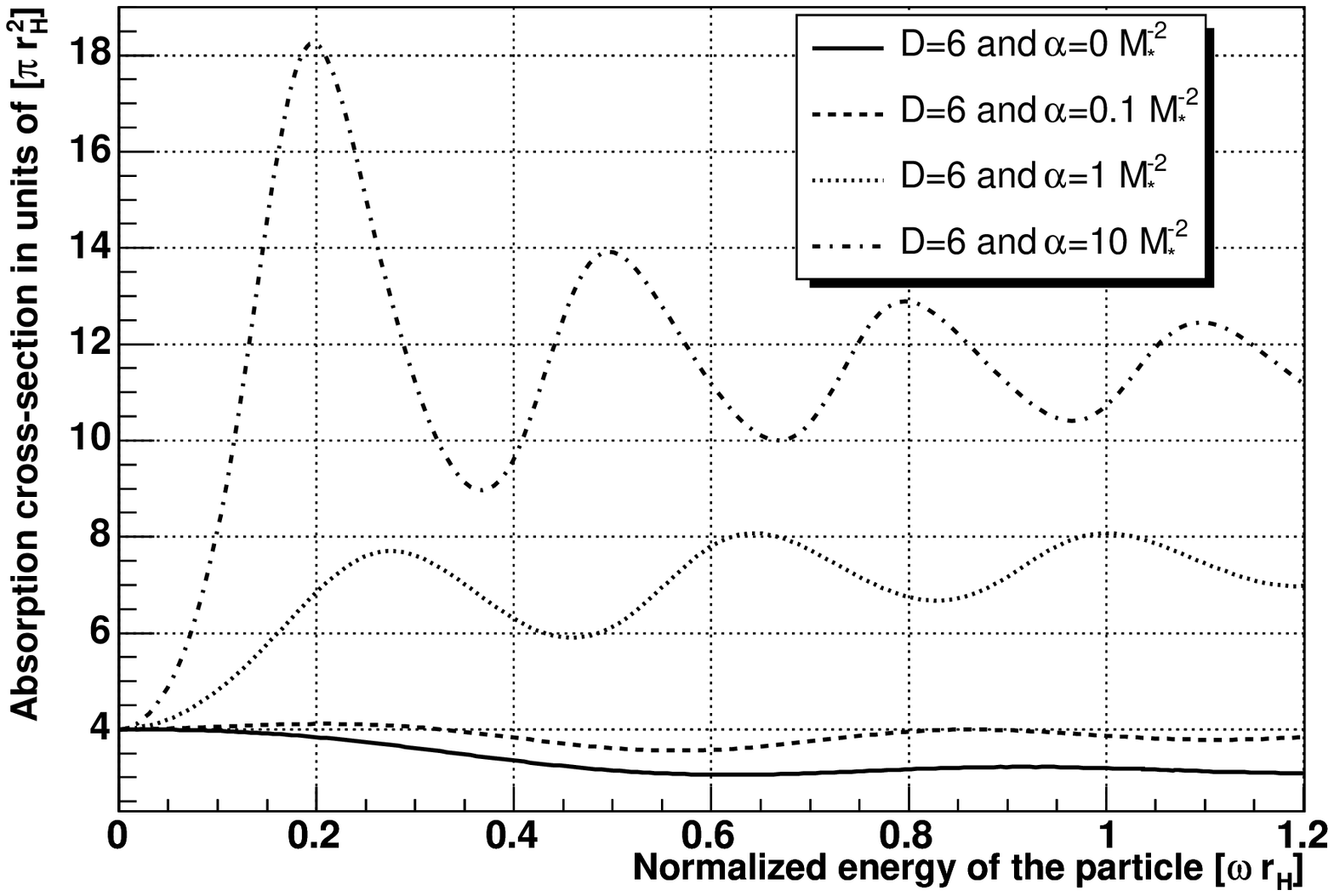}
		\end{tabular}
		\caption{Absorption cross-section for scalar fields on the brane for 
		 Schwarzschild-Gauss-Bonnet black holes as a function of the energy of 
		 the particle $\omega{r}_H$. {\it Left Panel~:} the Gauss-Bonnet 
		 coupling constant $\alpha$ is fixed at $0.1~M^{-2}_{*}$ and the number 
		 of dimensions $D$ at $\{5,6,10\}$. {\it Right Panel~:} the Gauss-Bonnet 
		 coupling constant takes the values $\{0,0.1,1,10\}~M^{-2}_{*}$ and the 
		 number of dimensions is fixed at $6$.}
		\label{scal-gre-brane}
	\end{center}
\end{figure}

Figure \ref{scal-gre-brane} shows the numerical results we have obtained for the 
absorption cross-section in units of $\pi{r}^2_H$ as a function of the dimensionless 
parameter $\omega{r}_H$. The two panels depict results for\,: a fixed value of the
Gauss-Bonnet coupling constant and variable $D$, the first one, and a fixed value of $D$ and
variable $\alpha$, the second. In both panels, when $\omega{r}_H \to 0$, the absorption 
cross-section tends to a non-vanishing value ($4\pi{r}^2_H$), which does not depend on 
the number of dimensions or on the value of the Gauss-Bonnet coupling constant. 
This behaviour follows the same pattern seen in other cases of asymptotically-flat
spherically-symmetric black hole backgrounds \cite{kanti2, harris, Jung-scalar} emitting
scalar radiation: in the low-energy regime, the absorption cross-section reduces to 
the area of the black hole horizon - for brane emission, this is simply $4\pi{r}^2_H$.

On the contrary, the behaviour of the cross-section in all other energy regimes strongly
depends on both the number of dimensions and on the Gauss-Bonnet coupling constant\,:
as for the case of Schwarzschild and Schwarzschild-de Sitter black holes, an increase
in the number of dimensions leads to a decrease of the absorption cross-section
\cite{harris,kgb}. On the other hand, an increase of the Gauss-Bonnet coupling constant
leads to an increase of the absorption cross-section. Both of these results could be
inferred from the behavior of the gravitational potential, that was derived in section 3.1,
with respect to $D$ and $\alpha$\,: the potential increases when $D$ increases 
whereas it decreases when $\alpha$ increases. The oscillatory behavior of the absorption
cross-section, that follows as modes with higher angular momentum numbers gradually come into
dominance, is also affected by the presence of the Gauss-Bonnet term\,: the frequency 
of the oscillations increases with the Gauss-Bonnet coupling constant.

\begin{figure}
	\begin{center}
		\begin{tabular}{cc}
			\includegraphics[scale=0.35]{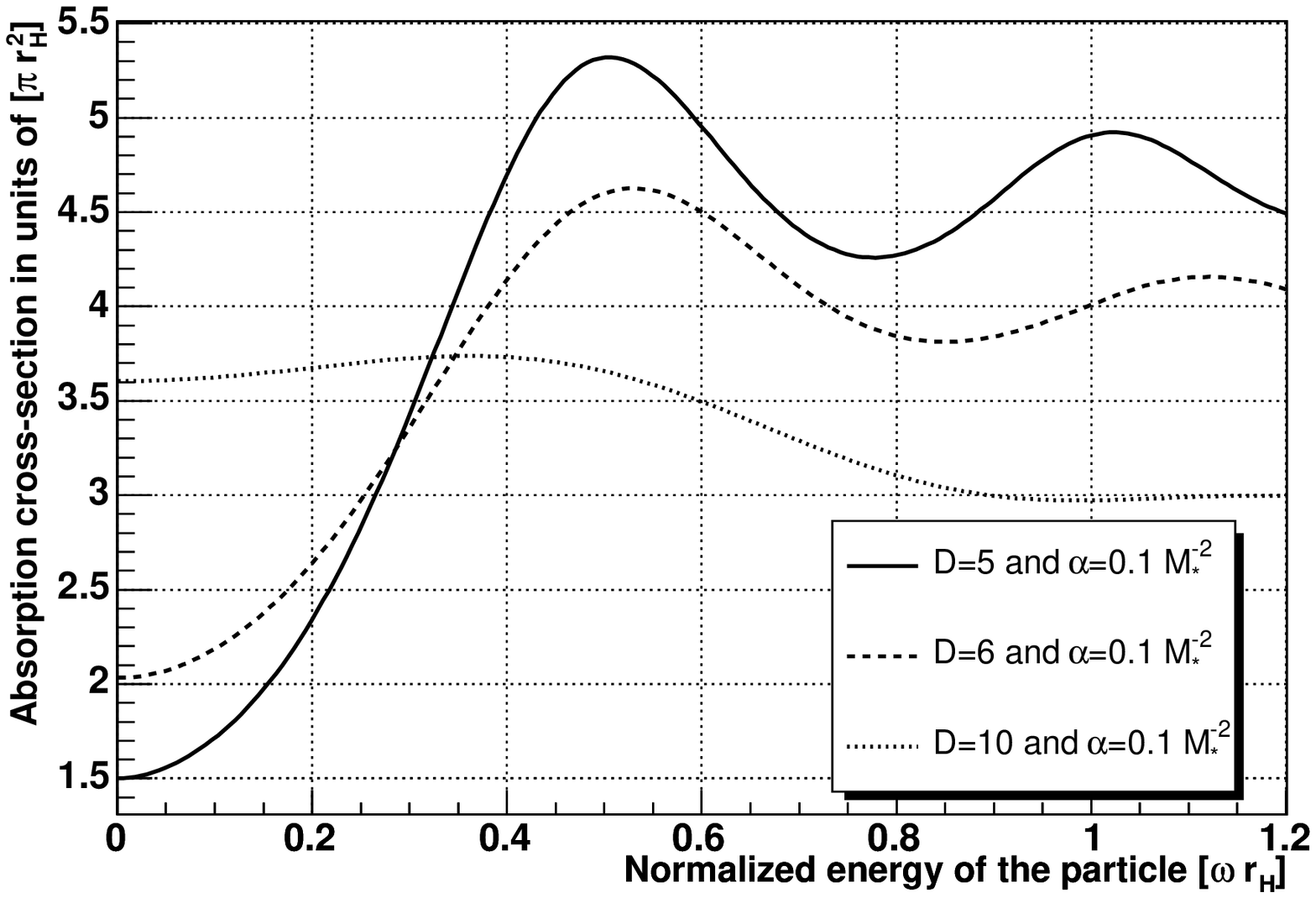}&
  \hspace*{-0.5cm}\includegraphics[scale=0.35]{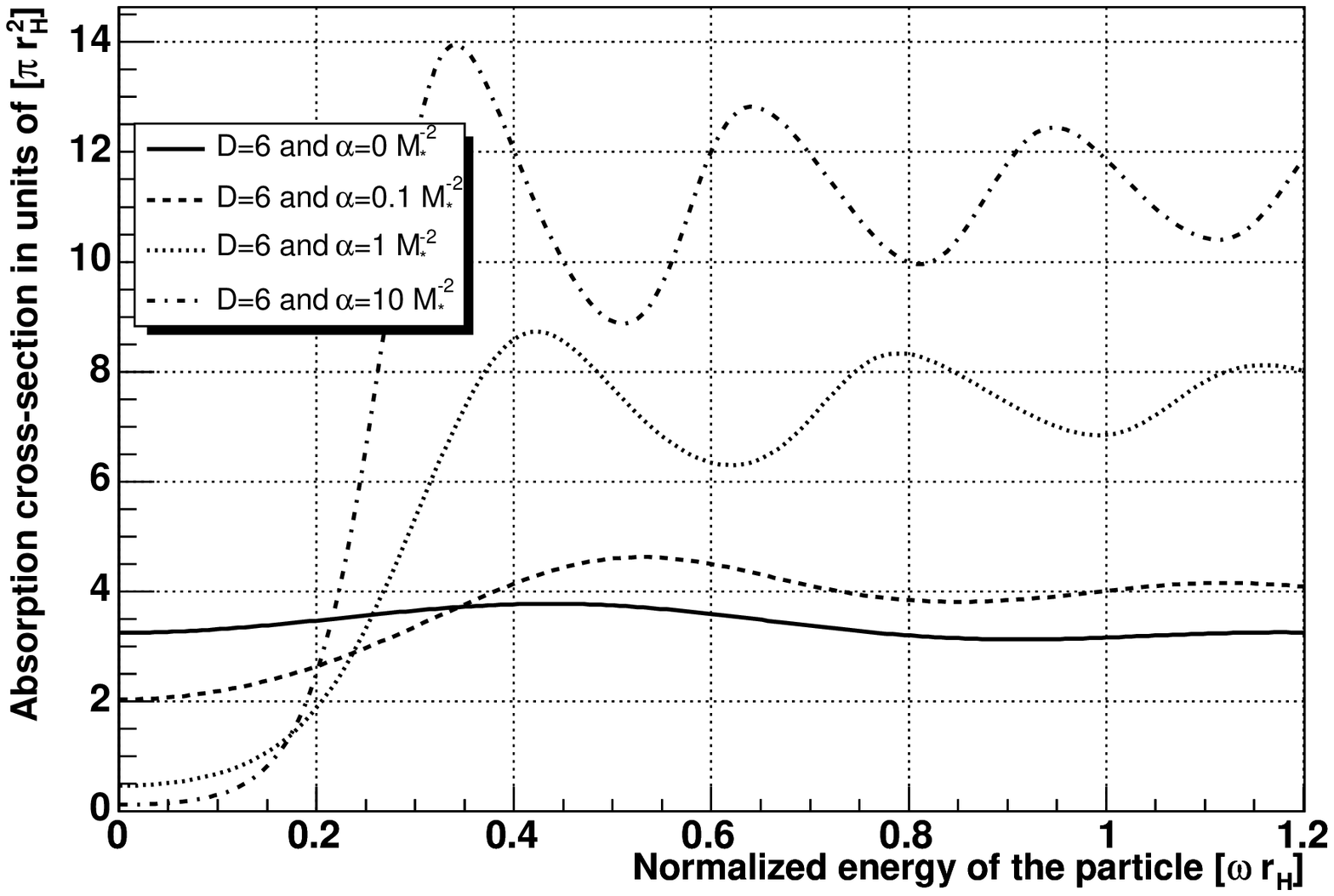}
		\end{tabular}
		\caption{Absorption cross-section for fermion fields on the brane for 
		 Schwarzs\-child-Gauss-Bonnet black holes as a function of the energy of
		 the particle $\omega{r}_H$. {\it Left Panel~:} the Gauss-Bonnet coupling
		 constant $\alpha$ is fixed at $0.1~M^{-2}_{*}$ and the number of 
		 dimensions $D$ at $\{5,6,10\}$. {\it Right Panel~:} the Gauss-Bonnet 
		 coupling constant takes the values $\{0,0.1,1,10\}~M^{-2}_{*}$ and $D$
                 is fixed at $6$.}
		\label{ferm-gre-brane}
	\end{center}
\end{figure}

\subsubsection{Absorption cross-section for fermions}

For fermion fields, the ingoing flux is defined as the radial component of the 
conserved current $J^\mu=2\sigma^\mu_{AB}\Psi_{1/2}^A\bar{\Psi}_{1/2}^B$ 
evaluated over a two dimensional sphere (see \cite{cvetic} for an explicit calculation). 
As in the case of scalar fields, this flux has to be calculated both 
at the black-hole event horizon and at spatial infinity in order to find 
$\left|\mathcal{A}_j\right|^2$. Once this is done, the absorption probability
for a mode with an angular momentum $j$ is found to be
\begin{equation}
	\left|\mathcal{A}_j\right|^2=\left|\frac{1}{B_1}\right|^2.
\end{equation}
Unlike the scalar case, we can not use the outgoing flux to determine the absorption
coefficient because we have only considered the upper component of the fermion fields.

Figure \ref{ferm-gre-brane} displays the numerical evaluation of the absorption 
cross-section for fermion particles on the brane. The high-energy limit of this 
cross-section is equal to the one for scalar particles and behaves in the same way with 
respect to the number of dimensions and to the Gauss-Bonnet coupling constant\,: the
high-energy limit increases with the Gauss-Bonnet coupling constant whereas it decreases with 
the number of dimensions. When $\omega{r}_H\to0$, the absorption cross-section tends to a 
non-vanishing value which behaves similarly to the Schwarzs\-child case with respect to the 
number of dimensions\,: an increase in the value of $D$ leads to an increase of the
low-energy asymptotic value of the absorption cross-section (we refer the reader to 
\cite{kanti2,harris} for an exhaustive discussion). The Gauss-Bonnet 
term leaves also a footprint on the asymptotic value of the cross-section when 
$\omega{r}_H\to0$~: this asymptotic value decreases when the Gauss-Bonnet coupling 
increases. As in the high energy limit, the Gauss-Bonnet coupling constant changes the 
absorption cross-section in an opposite way than the number of dimensions does.

\subsubsection{Absorption cross-section for gauge bosons}

For gauge boson fields, the ingoing flux is derived from the trace of the
energy-momentum tensor $T^{\mu\nu}=2\sigma^\mu_{AB}\sigma^\nu_{A'B'}\Psi_{1}^{AB}
\bar{\Psi}_{1}^{A'B'}$  evaluated, as for fermions, over a two dimensional sphere.
An explicit calculation of the flux at the black hole's event horizon and at spatial
infinity leads to the absorption coefficient \cite{cvetic,kanti3}:
\begin{equation}
	\left|\mathcal{A}_j\right|^2=\frac{1}{r_H^2}\left|\frac{1}{B_1}\right|^2\,.
\end{equation}

\begin{figure}
	\begin{center}
		\begin{tabular}{cc}
			\includegraphics[scale=0.35]{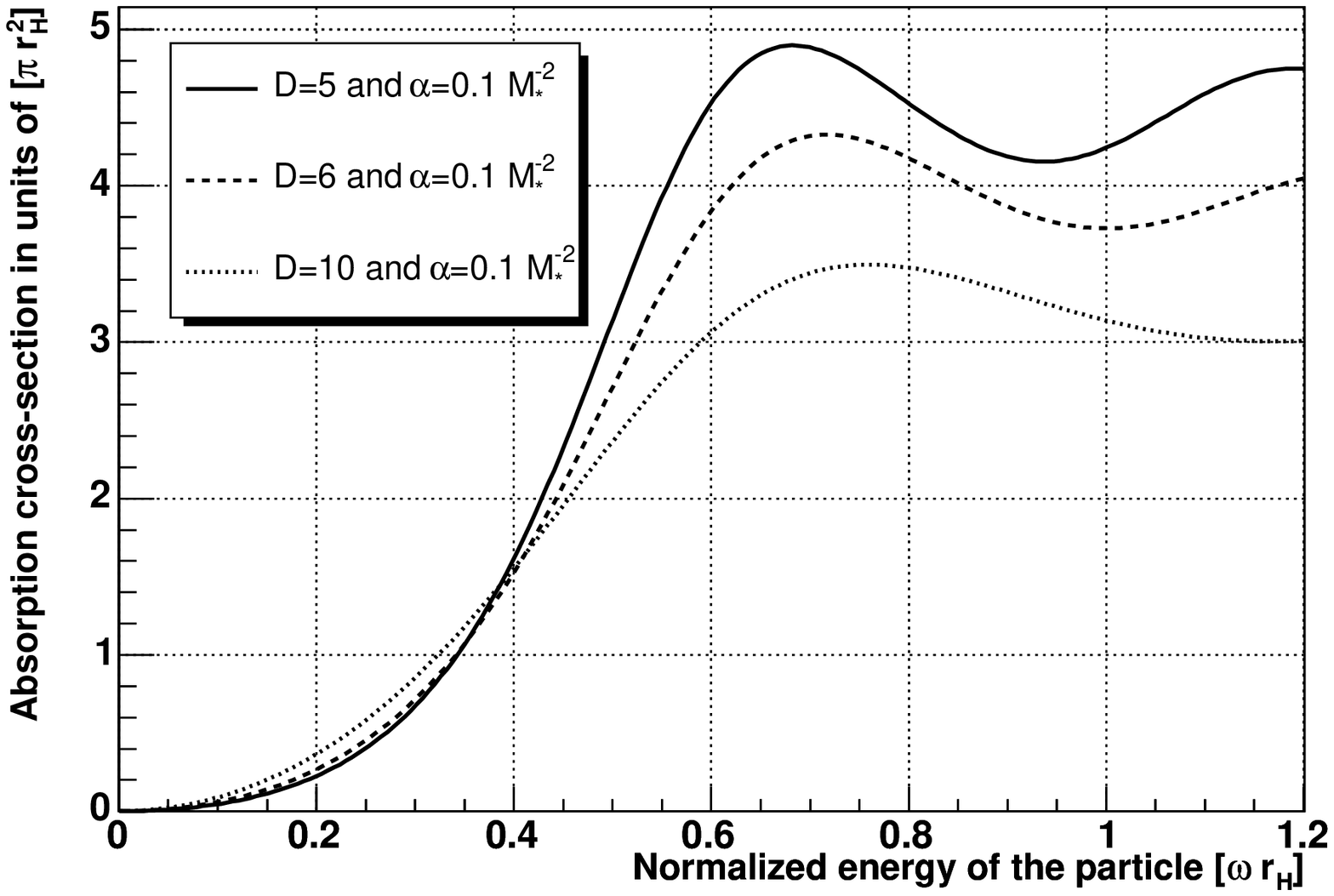}&
 \hspace*{-0.5cm} \includegraphics[scale=0.35]{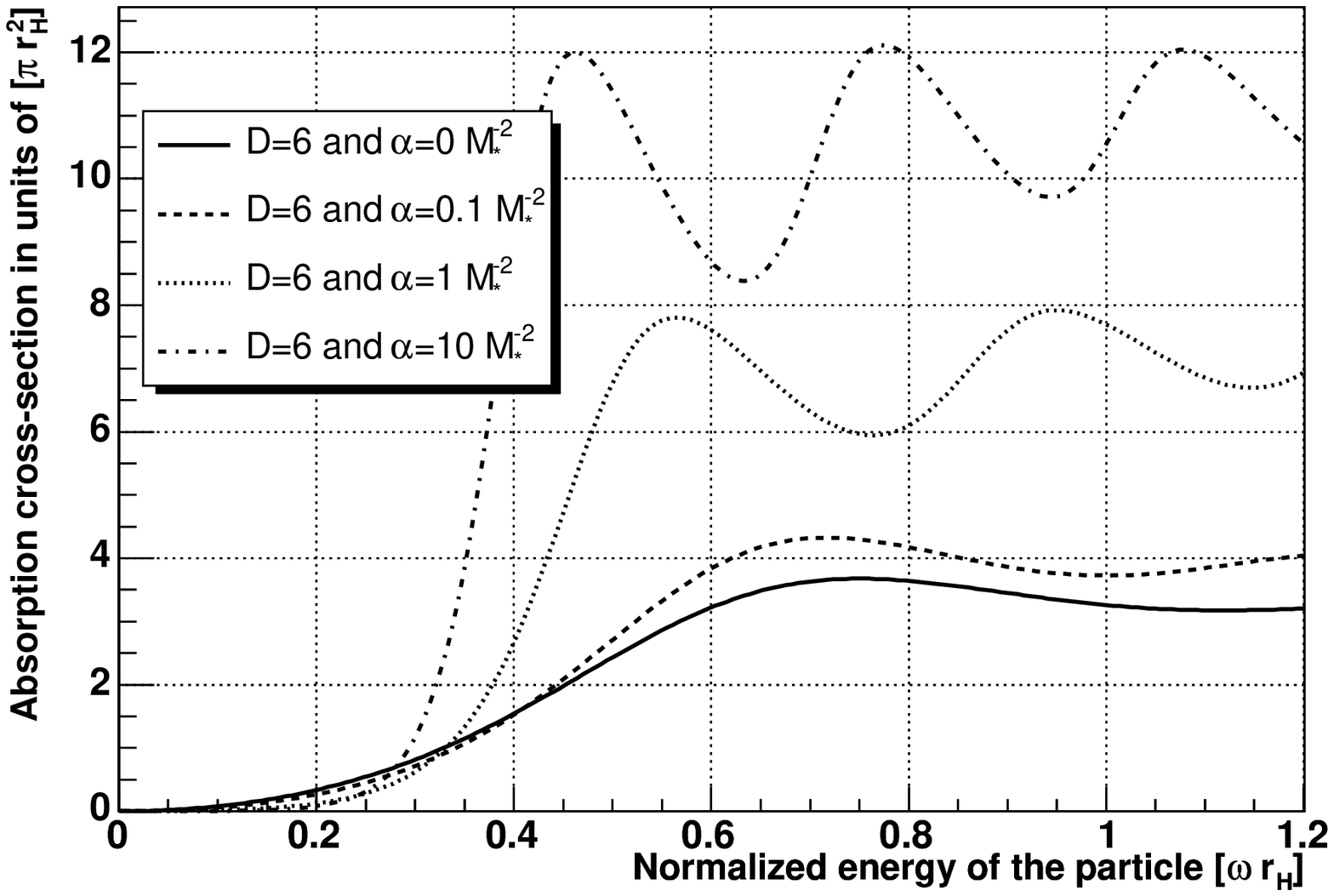}
		\end{tabular}
		\caption{Absorption cross-section for gauge bosons on the brane 
		 for Schwarzs\-child-Gauss-Bonnet black holes as a function of the 
		 energy of the particle $\omega{r}_H$. {\it Left Panel~:} the Gauss-Bonnet
		 coupling constant $\alpha$ is fixed at $0.1~M^{-2}_{*}$ and the number 
		 of dimensions $D$ at $\{5,6,10\}$. {\it Right Panel~:} the Gauss-Bonnet 
		 coupling constant takes the values $\{0,0.1,1,10\}~M^{-2}_{*}$ and $D$
		 is fixed at $6$.}
		\label{bose-gre-brane}
	\end{center}
\end{figure}

Despite a few quantitative differences, we may clearly see from Fig. \ref{bose-gre-brane}
that the absorption cross-section in the case of gauge bosons exhibits a similar behavior  
to fermions with respect to the number of dimensions and to the Gauss-Bonnet coupling 
constant. The high-energy regime asymptotic values are again suppressed with the number
of dimensions and enhanced with the value of the Gauss-Bonnet coupling constant. The
main difference occurs in the asymptotic regime of low energy\,: the absorption 
cross-section tends to zero when $\omega{r}_H\to0$, a behavior exhibited also in
the pure Schwarzschild case \cite{kanti3}. Although the asymptotic value of the
absorption cross-section vanishes in this regime, the statement that an increase in
the Gauss-Bonnet coupling constant leads to a decrease of the cross-section at low
energy is still valid as it falls down to zero more rapidly for higher values of 
$\alpha$. On the contrary, the cross-section approaches zero slower for higher 
values of the number of dimensions.

From the previous results, we can summarize the effect of the Gauss-Bonnet term on the 
absorption cross-section as follows~: for scalar particles, an increase of the Gauss-Bonnet
coupling constant leads to an increase of the absorption cross-section over the whole 
energy spectrum. For particles with non-vanishing spin, an increase of the Gauss-Bonnet 
coupling constant leads to an increase of the absorption cross-section in the high-energy
regime whereas it decreases it in the low-energy regime. Finally, the effect of the 
number of dimensions is the same as for Schwarzschild black holes and affects the 
cross-section in exactly the opposite way that the Gauss-Bonnet coupling constant does.
As a similar behavior is also exhibited by the temperature of the black hole in
terms of its dependence on $\alpha$ and $D$, it becomes increasingly more interesting
to compute the final emission rate to decide on the prevailing factor (temperature or
absorption cross-section) that shapes the radiation spectrum for this particular
black-hole background.

\subsection{Semi-analytical results for the high energy limit}

As it is known, and was numerically checked in the previous subsections, in the
high-energy regime, the greybody factor assumes its geometrical optics limit value,
which is independent of the spin of the emitted particle. This limiting value has
been successfully used in both 4-dimensional and higher-dimensional black-hole
backgrounds \cite{harris, kgb} to describe the greybody factor of the corresponding
black hole. This technique makes use of the geometry around the emitting (or absorbing)
body and the same method can be applied for a Schwarzschild-Gauss-Bonnet projected
black hole: for a massless particle in a circular orbit around a black hole, described
by a line-element of the form (\ref{metric-brane}), its equation of motion, 
$p^\mu p_\mu=0$, can be written in the form
\begin{equation}
\biggl(\frac{1}{r}\,\frac{dr}{d\varphi}\biggr)^2=\frac{1}{b^2}
-\frac{h(r)}{r^2}\,,
\label{circular}
\end{equation}
where $b$ is the ratio of the angular momentum of the particle over its linear
momentum. The classically accessible regime is defined by the relation
$b< {\rm min}(r/\sqrt{h})$. The above result is valid for all spherically-symmetric
Schwarzschild-like line-elements of the form (\ref{metric-brane}) with the structure
of the particular gravitational background entering the above relation through
the metric function $h(r)$. In order to compute the critical value of the parameter
$b$, we must first minimize the function $r/\sqrt{h}$; this takes place at $r=r_c$,
where $r_c$ is defined through the relation
\begin{equation}
\sqrt{1+\frac{64\pi \alpha\,G M_{BH}\,(D-3)(D-4)}{(D-2)
\,\Omega_{D-2}\,r_c^{D-1}}}=\frac{8 \pi (D-1) GM_{BH}}{(D-2)\,\Omega_{D-2}\,r_c^{D-3}}\,. 
\label{rc}
\end{equation}
Then, $b_c$ is found by evaluating the function $r/\sqrt{h(r)}$ at the value $r=r_c$.
The corresponding area, $\sigma_g = \pi b_c^2$, defines the absorptive area of the
black hole at high energies and, thus, its greybody factor -- being a constant, the
emitting body is characterized at high energies by a blackbody radiation spectrum.
Its exact expression comes out to be
\begin{equation}
\sigma_g = \frac{\pi r_c^2}{1+\frac{\textstyle r_c^2}{\textstyle 2\alpha(D-3)(D-4)}
\left(1-\sqrt{1+\frac{\textstyle 64\pi \alpha\,G M_{BH}\,(D-3)(D-4)}{\textstyle (D-2)
\,\Omega_{D-2}\,r_c^{D-1}}}\right)}\,.
\label{optics}
\end{equation}
For $D=4$ and $\alpha=0$, $r_c=3 r_H$ and the above expression reduces to the
four-dimensional one, $\sigma_g= 27 \pi r_H^2/4$. For arbitrary values, however, of 
$D$ and $\alpha$, Eq. (\ref{rc}) can not be solved in a closed form, and the
value of $r_c$ can be found only numerically in terms of $D$, $\alpha$, and, through
Eq. (\ref{Mass}), $r_H$. In Table 1, we display the values of the greybody factor
under the geometric optics approximation, in units of $\pi r_H^2$, for various values
of $D$ and $\alpha$, and for $r_H=1$. It can easily been checked that these results
are in excellent agreement with the high-energy behavior of the numerically computed
greybody factors presented in the previous subsections.

\begin{table}[t]
\begin{center}
\begin{tabular}{|p{1.8cm}|*{5}{c|}|}
\hline
$\sigma_g/(\pi r_H^2)$ & $\alpha=0$ & $\alpha=0.1$ & $\alpha=1$  & $\alpha=10$\\
\hline                       
D=5 & 4.00 & 4.59 & 9.46 & 51.2\\
\hline
D=6 & 3.07 & 3.91 & 7.64 & 11.4\\
\hline                       
D=7 & 2.60 & 3.53 & 5.70 & 6.62\\
\hline                       
D=8 & 2.31 & 3.24 & 4.53 & 4.88\\
\hline
D=9 & 2.12 & 3.00 & 3.82 & 3.98\\
\hline
D=10 & 1.98 & 2.80 & 3.34 & 3.43\\
\hline
D=11 & 1.87 & 2.63 &  3.00 & 3.06\\
\hline
\end{tabular}
\end{center}
\caption{
High-energy limits of the greybody factor on the brane in units of $\pi r_H^2$.}
\label{chi2_tab}
\end{table}

\section{Radiation in the bulk}

\subsection{Equation for scalar particles in the bulk}

Although all Standard Model (SM) particles are confined to our $4-$dimensional brane, 
scalar particles which are neutral with respect to the SM charges are allowed to propagate 
in the bulk. As a consequence, black holes can radiate spin-0 particles out of the brane
and this radiation, although invisible to us, contributes to the total mass-loss rate.
In order to make accurate estimates of the amount of energy radiated by the black hole
on the brane, it is therefore necessary to determine the absorption cross-section
and eventually the energy emission rate for those particles, too.

The generalization of the Klein-Gordon equation to a $D-$dimensional curved space-time is 
given by $\Psi^{;\mu}_{~;\mu}=0$, where $\mu$ runs over the $D$ dimensions. In the 
spherically symmetric case, the following ansatz is used~:
\begin{equation}
	\Psi(t,r,\theta_i,\varphi)=e^{-i\omega{t}}P(r)\,\tilde{Y}^m_\ell(\theta_i,\varphi)
\end{equation}
where $\tilde{Y}^m_{\ell}(\theta_i,\varphi)$ stands for the $D-$dimensional generalization 
of the spherical harmonics \cite{muller}. With this ansatz, the radial part of the 
Klein-Gordon equation, satisfied by the unknown function $P(r)$, takes the form\,:
\begin{equation}
\frac{h(r)}{r^{D-2}}\,\frac{d}{dr}\left(r^{D-2}h(r)\,\frac{dP}{dr}\right)+
\left[\omega^2-\frac{h(r)}{r^2}\left(\ell+D-3\right)\right]P=0.
	\label{bulk-eq}
\end{equation}
Apart from an implicit dependence upon the number of dimensions through the metric
function $h$, the bulk equation exhibits also an explicit dependence through the
contribution of the $D-2$ compact dimensions entering the kinetic term and through the 
eigenvalues associated to the $D-$dimensional spherical harmonics. These contributions
modify the exact form of the centrifugal potential, however, its general behaviour
remains the same vanishing both at the black hole horizon and spatial infinity. Therefore
the solution of the above equation at these two asymptotic regimes will again be given
in terms of plane waves. 

Although the basis for the Hawking radiation process in the bulk is the same as for
emission on the brane, both analytical and numerical studies of Eq. (\ref{bulk-eq})
have shown that varying the number of dimensions have different quantitative consequences
on the bulk greybody factor and emission rate, for Schwarzschild \cite{kanti2,harris}
and Schwarzschild-de Sitter black holes \cite{kgb}. As in those studies, we expect the
number of dimensions as well as the Gauss-Bonnet coupling constant to strongly modify
the value of the absorption cross-section for scalar particles in the case of a 
Schwarzschild-Gauss-Bonnet black hole, too.

\subsection{Numerical results for the absorption cross-section}
	
We numerically evaluate the absorption cross-section for emission in the bulk by
following the same method as in the case of brane emission\,: Eq. (\ref{bulk-eq}) is
numerically integrated starting from the vicinity of the black hole's event horizon,
at $r_{ini}=r_H+\varepsilon$, where the boundary conditions (\ref{bound1})-(\ref{bound2})
have been again applied, and propagating the solution until spatial infinity. The
analytical asymptotic solutions at spatial infinity, given by \cite{kanti2}
\begin{equation}
P_0(r)=B_1\,\frac{e^{-i \omega r}}{\sqrt{r^{D-2}}} + 
B_2\,\frac{e^{i \omega r}}{\sqrt{r^{D-2}}}\,,
\end{equation}
are then fitted on the numerical results and the amplitude of both ingoing and
outgoing modes are extracted from the fit parameters. Once again, the absorption
probability $\left|{\cal A}_\ell\right|^2$ is linked to these amplitudes via
Eq. (\ref{scal-coef}), and the absorption cross-section is obtained 
using the $D-$dimensional generalization of the optical theorem given in
Eq. (\ref{abs-cros}).

\begin{figure}
	\begin{center}
		\begin{tabular}{cc}
			\includegraphics[scale=0.35]{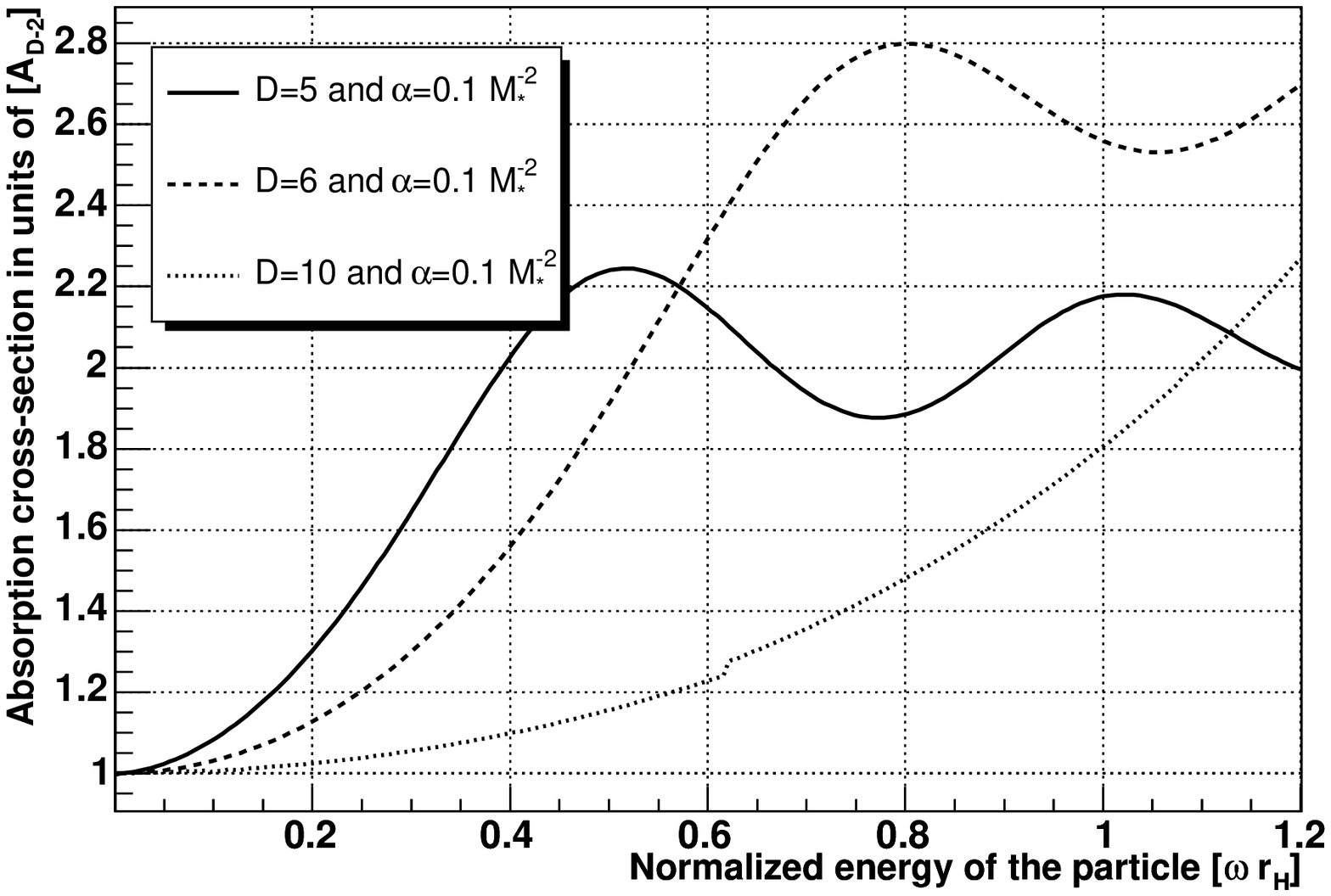}&
 \hspace*{-0.5cm} \includegraphics[scale=0.35]{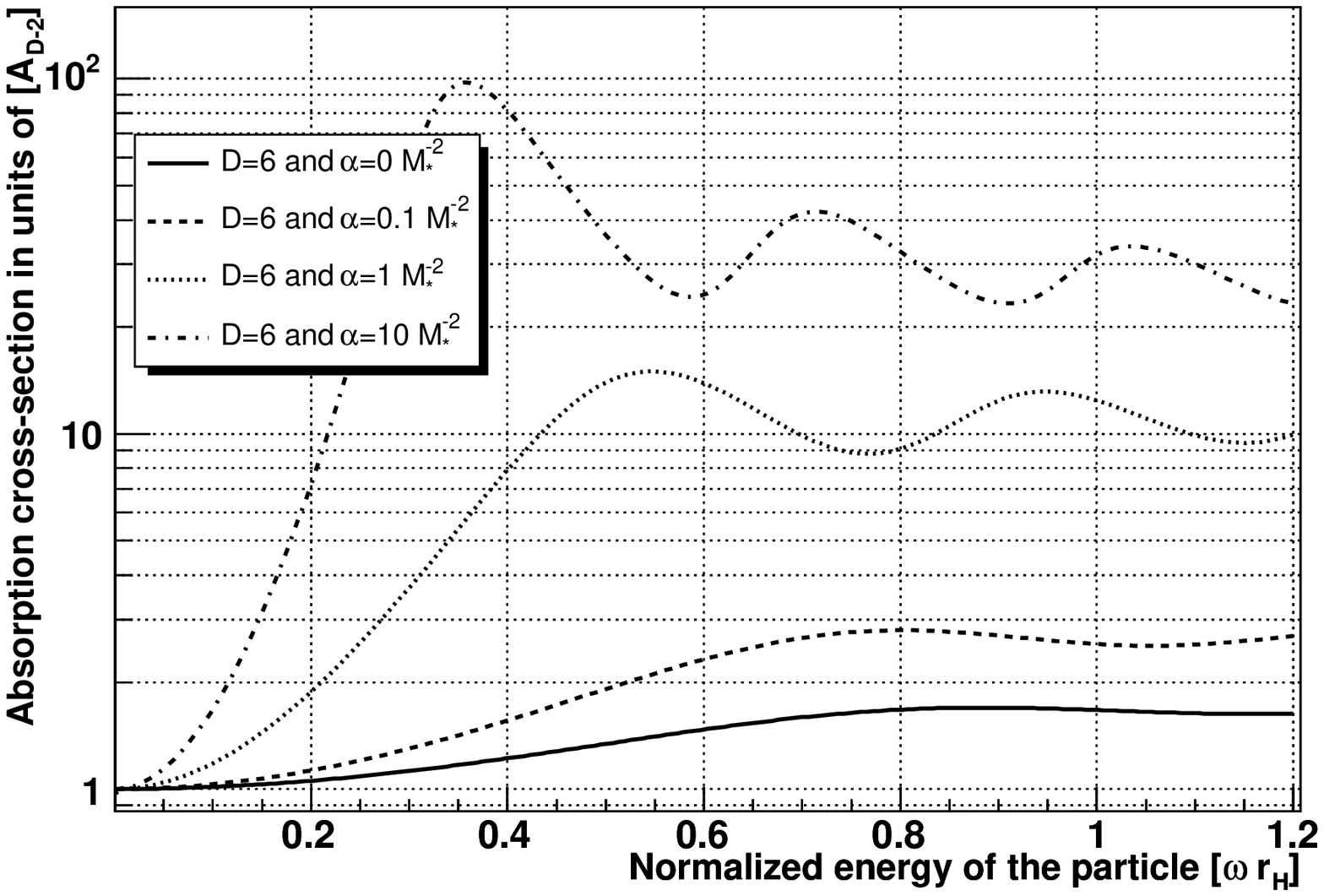}
		\end{tabular}
		\caption{Absorption cross-section for scalar fields in the bulk 
		for Schwarzschild-Gauss-Bonnet black holes as a function of the 
		energy of the particle $\omega{r}_H$. {\it Left Panel\,:} the 
		Gauss-Bonnet coupling constant $\alpha$ is fixed at 
		$0.1~M^{-2}_{*}$ and the number of dimensions $D$ at 
		$\{5,6,10\}$. {\it Right Panel\,:} the Gauss-Bonnet coupling 
		constant takes the values $\{0,0.1,1,10\}~M^{-2}_{*}$ and the number 
		of dimensions is fixed at $6$.}
		\label{scal-gre-bulk}
	\end{center}
\end{figure}

Our results are shown in Fig. \ref{scal-gre-bulk} where the absorption cross-section is 
expressed in units of the black hole's area $A_{D-2}$, {\it{i.e.}} the area of a 
$(D-2)-$dimensional sphere with radius $r_H$
\begin{equation}
A_{D-2}=r_H^{D-2}\,(2 \pi)\,\pi^{(D-3)/2}\,\Gamma 
\biggl(\frac{D-1}{2}\biggr)^{-1}\,.
\end{equation}
These plots exhibit the same qualitative behavior as for the brane emission case\,:
the absorption cross-section tends to a non-vanishing value when $\omega{r}_H\to0$,
increases with the Gauss-Bonnet coupling constant while it decreases for higher D values. 
From a quantitative point of view, the number of dimensions leaves a clear footprint on
the asymptotic value in the low energy regime as $\sigma \rightarrow {A}_{D-2}$, which
only depends on the size of the black hole and the number of dimensions, a behavior
similar to the one found in the $D-$dimensional Schwarzschild black hole case \cite{harris}.
The effect of the Gauss-Bonnet coupling constant, on the other hand, is greatly enhanced
when compared to what happens in the case of brane emission. The significant enhancement
of the bulk absorption cress-section, as the Gauss-Bonnet coupling constant increases,
will play a significant role in the determination of the bulk-to-brane emission rates
ratio to be computed in Section 5.3. Another novel characteristic, that arises in the
case of a Schwarzschild-Gauss-Bonnet black hole, is the fact that the bulk absorption
cross-section is enhanced by the number of extra dimensions from the intermediate
energy-regime and onwards - in the case of a pure $D-$dimensional Schwarzschild 
black hole, the absorption cross-section in the bulk was heavily suppressed with 
the number of dimensions in the low- and intermediate-energy regimes allowing only
for a small enhancement in the high-energy part of the spectrum \cite{harris, kanti1, kgb}.


\begin{table}[t]
\begin{center}
\begin{tabular}{|p{1.5cm}|*{5}{c|}|}
\hline
$\sigma_g/A_{D-2}$ & $\alpha=0$ & $\alpha=0.1$ & $\alpha=1$  & $\alpha=10$\\
\hline                       
D=5 & 1.70 & 2.09 & 6.19 & 77.7\\
\hline
D=6 & 1.77 & 2.87 & 11.0 & 24.5\\
\hline                       
D=7 & 1.85 & 3.97 & 13.2 & 19.1\\
\hline                       
D=8 & 1.93 & 5.31 & 14.6 & 18.1\\
\hline
D=9 & 2.01 & 6.81 & 15.8 & 18.3\\
\hline
D=10 & 2.08 & 8.40 & 17.0 & 19.0\\
\hline
D=11 & 2.16 & 10.0 & 18.3 & 19.9\\
\hline
\end{tabular}
\end{center}
\caption{High-energy limit of the greybody factor in the bulk in units of
$A_{D-2}$.}
\label{bulk-gb}
\end{table}

\subsection{Semi-analytical results for the high energy limit}

For emission in the bulk, the high-energy limit of the absorption cross-section is
again given by the geometric optics approximation. The classically accessible regime
for $D-$dimensional spacetimes of the form described in Eq. (\ref{metric-bulk}) is
again given by $b< {\rm min}(r/\sqrt{h})$ \cite{Emparan}, with the minimum radial
proximity $b_c$ obtained as in Section 3.3. What nevertheless changes, in the case
of propagation in a higher-dimensional black-hole background, is the expression
of the absorptive area of the black hole as a function of the critical radius
$b_c$\,: as it was pointed out in Ref. \cite{harris}, when carefully computed, this
area comes out to be
\begin{equation}
A_p = \frac{2 \pi}{(D-2)}\,\frac{\pi^{(D-4)/2}}{\Gamma[(D-2)/2]}\,
b_c^{\,D-2}\,.
\end{equation}
This area stands for the absorption cross-section $\sigma_g$, or greybody factor,
for particle emission in the bulk at high energies.  

In Table \ref{bulk-gb}, we give some indicative values of the greybody factor, in units
of the area of the $D$-dimensional horizon $A_{D-2}$, for emission of scalar fields in
the bulk at high-energies, for particular values of $D$ and $\alpha$, and for
$r_H=1$. Once again, these asymptotic values are in good agreement with the numerical 
results obtained in the $\omega r_H \to \infty$ limit, and reveal the enhancement
of the absorption cross-section in this energy regime, in terms of both the
Gauss-Bonnet coupling constant and the dimensionality of spacetime.

\section{Particle emission spectrum}
	
In this section, the Hawking radiation spectrum emitted by a Schwarzschild-Gauss-Bonnet
black hole is computed using the exact numerical results derived in the previous
sections for the absorption cross-section. The spectrum is evaluated for scalar
particles, emitted on the brane and in the bulk, and for fermion and gauge boson
particles emitted on the brane. 
We also compare the number of scalar particles emitted on the brane and in the 
bulk as a function of the Gauss-Bonnet coupling constant, number of dimensions
and mass of the evaporating black hole. It was pointed in 
Ref. \cite{kgb} that the calculation of the temperature of a Schwarzschild-de 
Sitter black hole is not simply given by the surface gravity as it is not 
possible to place an observer at spatial infinity. Such a modification in the 
temperature computation is a direct consequence of the intrinsic curvature of 
space-time. In the present case, the space-time is asymptotically flat and 
we can safely use the surface gravity to compute the Schwarzschild-Gauss-Bonnet
black hole temperature given by equation (\ref{temp}). Consequently, the 
radiation spectra derived hereafter are relative to an observer infinitely far
from the black hole.

Up to this point, all thermodynamical quantities of the Schwarzschild-Gauss-Bonnet
black hole, as well as the greybody factors, were conveniently expressed in terms of
the event horizon radius $r_H$. Comparing the values of these quantities, as the
parameters $D$ and $\alpha$ varied while $r_H$ remained fixed, was indeed an
easy task. It also allowed for a direct comparison of our results to 
those derived previously in the literature under the same normalization. However,
from an observational point of view, the horizon radius is not a convenient
parameter to measure (and thus to fix), as different values of $D$ and $\alpha$
correspond to black holes with different masses. Therefore, for observational
convenience, in what follows, the various radiation spectra will be derived and
compared at fixed black hole masses.

\subsection{Radiation spectrum on the brane}

\begin{figure}
	\begin{center}
		\begin{tabular}{cc}
			\includegraphics[scale=0.35]{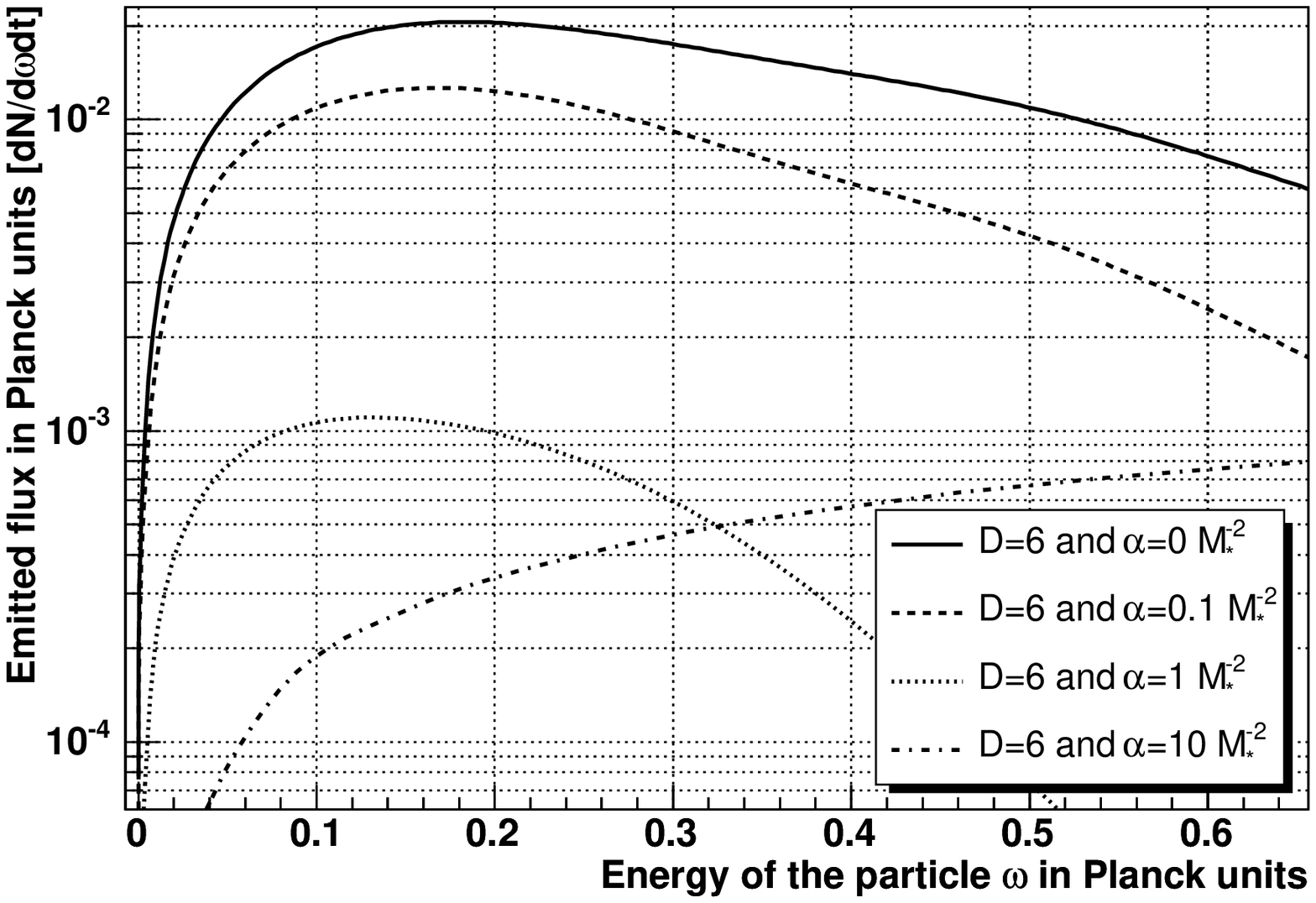}&
             \hspace*{-0.5cm} \includegraphics[scale=0.35]{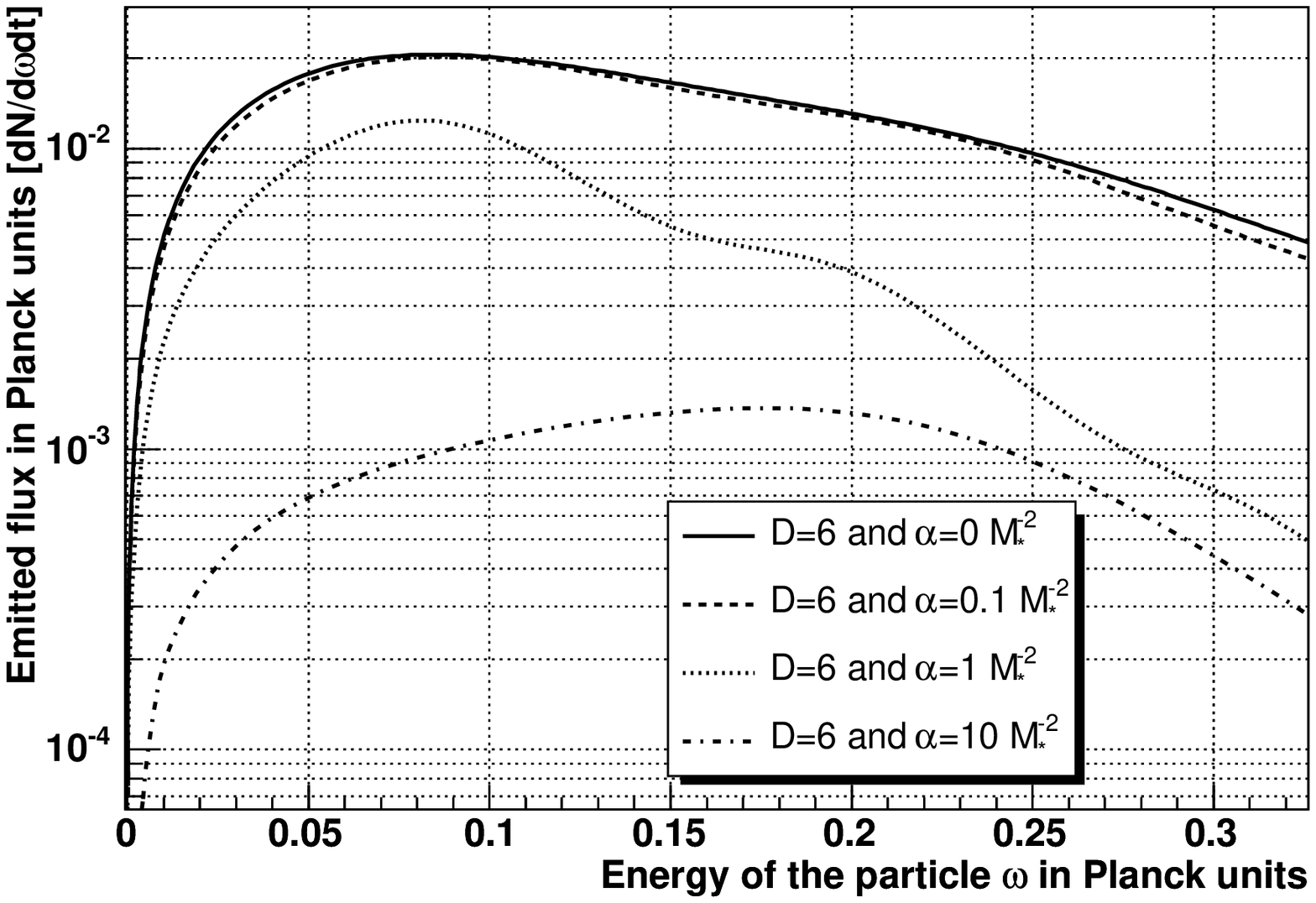} \\
			\includegraphics[scale=0.35]{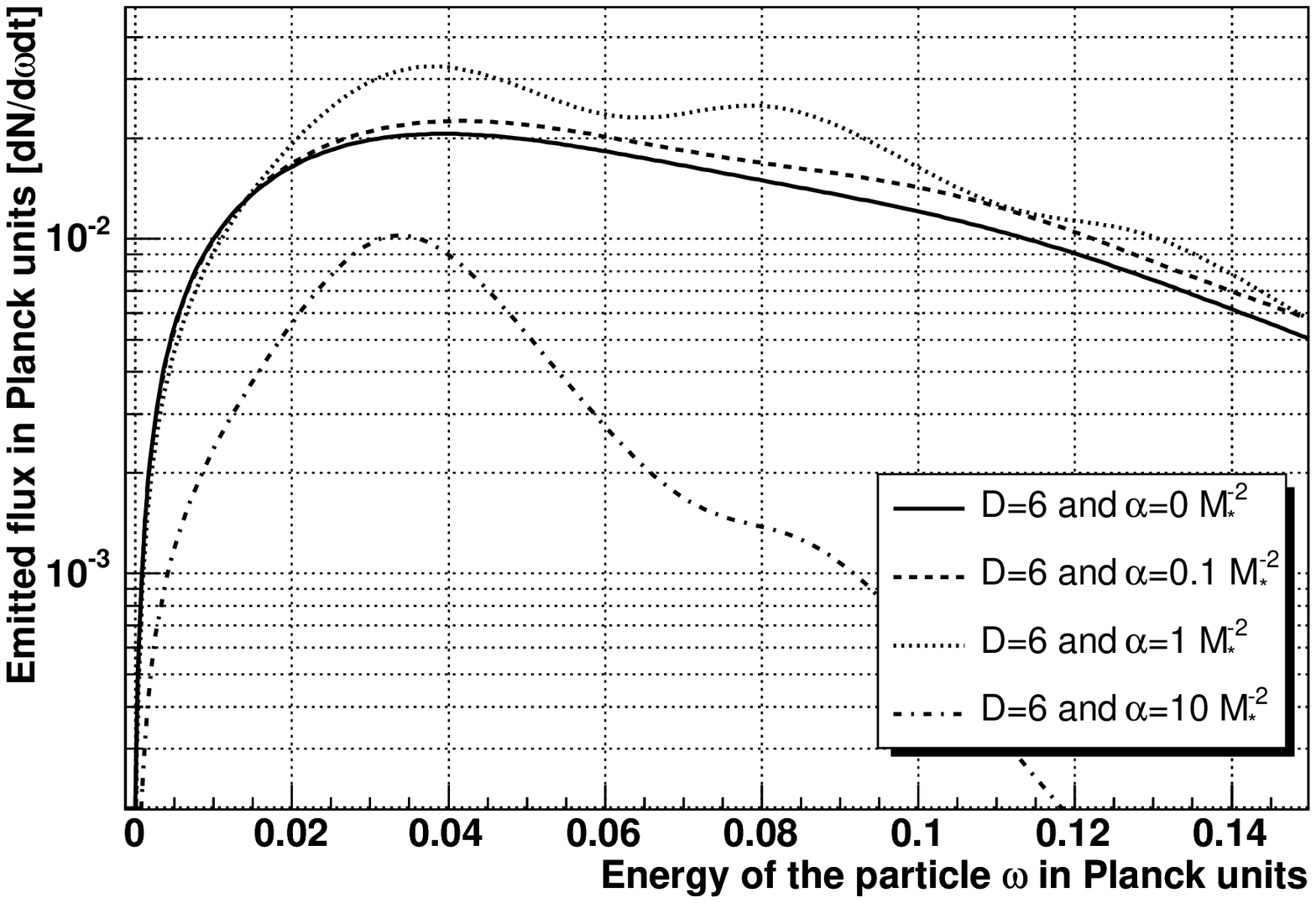}&
             \hspace*{-0.5cm} \includegraphics[scale=0.35]{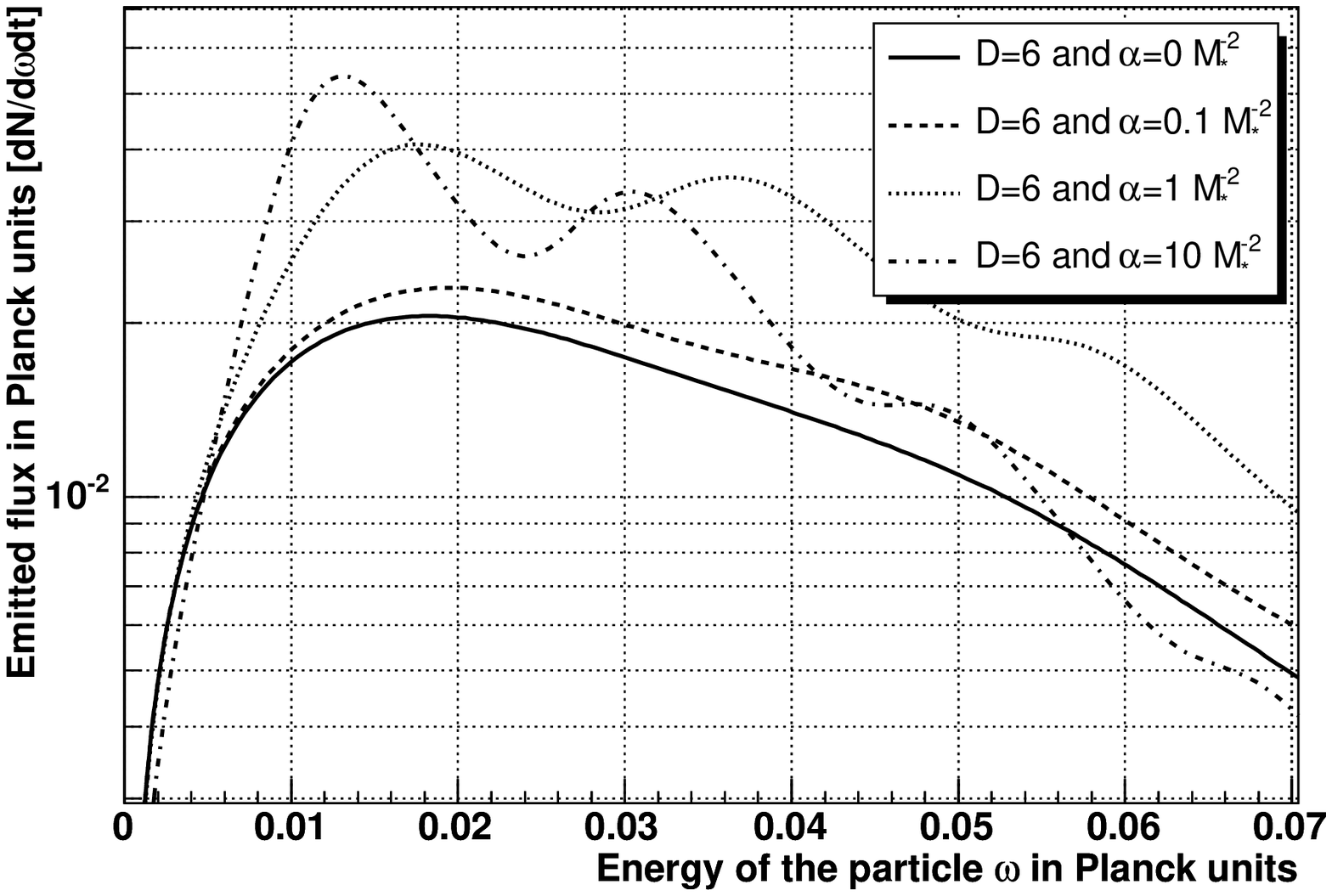}
		\end{tabular}
		\caption{Flux spectrum for scalar fields emitted on the brane 
		by a Schwarzschild-Gauss-Bonnet black hole as a function of the 
		energy $\omega$. The Gauss-Bonnet coupling constant
		takes the values $\{0,0.1,1,10\}\,M^{-2}_{*}$ and the number of 
		dimensions is fixed at $6$. The mass of the black hole is taken equal to
		 $M_{BH}=10~M_{*}$ (upper left), $M_{BH}=100~M_{*}$ 
		 (upper right), $M_{BH}=1000~M_{*}$ (lower left) and 
		 $M_{BH}=10000~M_{*}$ (lower right).}
		\label{scal-flu-brane}
	\end{center}
\end{figure}

\begin{figure}
	\begin{center}
		\begin{tabular}{cc}
			\includegraphics[scale=0.35]{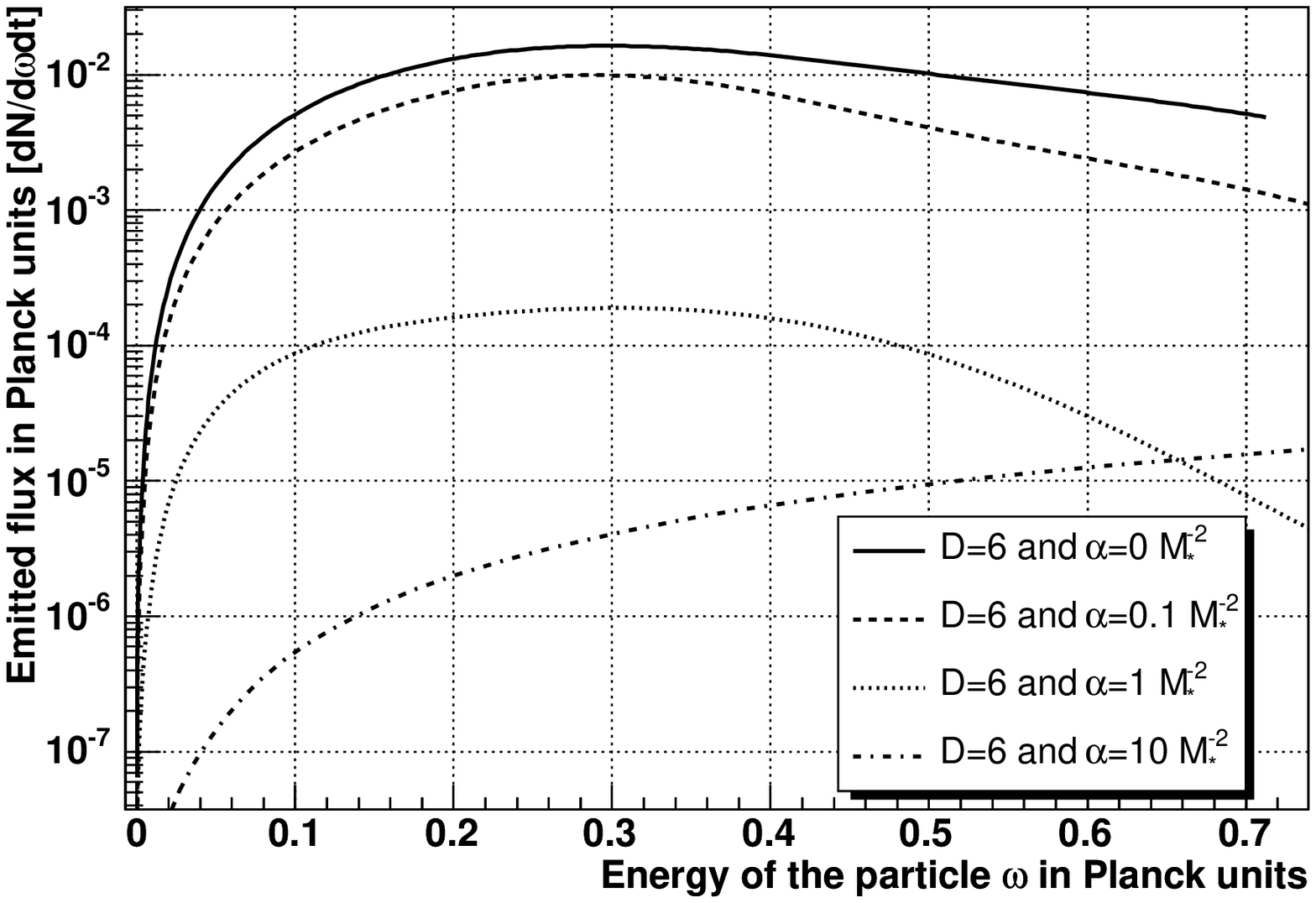}&
 \hspace*{-0.5cm} \includegraphics[scale=0.35]{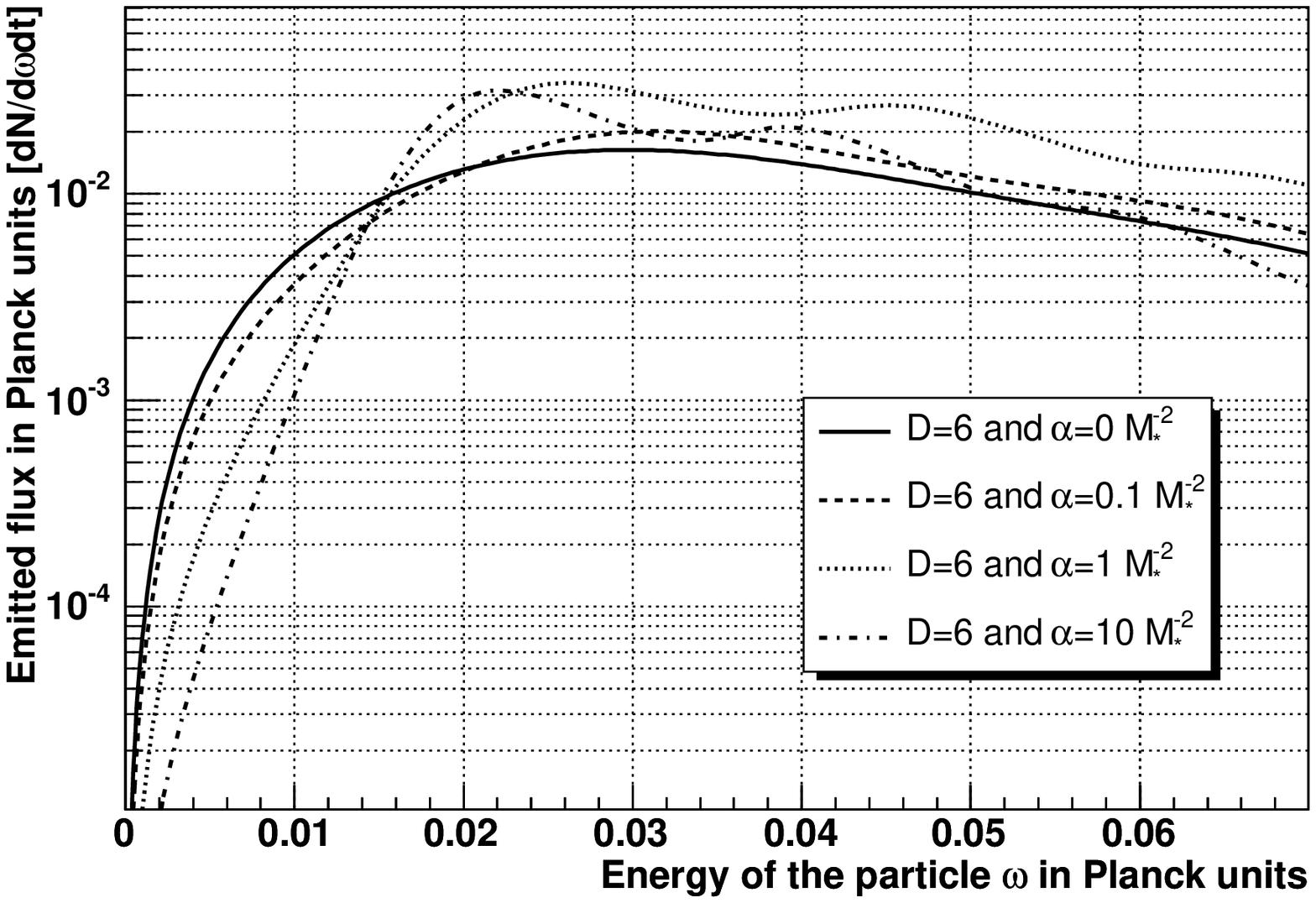}
		\end{tabular}
		\caption{Flux spectrum for fermion fields emitted on the brane 
		by a Schwarzs\-child-Gauss-Bonnet black hole as a function of the 
		energy $\omega$. The Gauss-Bonnet coupling 
		constant takes the values $\{0,0.1,1,10\}\,M^{-2}_{*}$ and the number 
		of dimensions is fixed at $6$. The mass of the black hole is 
                $M_{BH}=10\,M_{*}$ on the left panel, and $M_{BH}=10000\,M_{*}$
                 on the right panel.}
		\label{ferm-flu-brane}
	\end{center}
\end{figure}

\begin{figure}
	\begin{center}
		\begin{tabular}{cc}
			\includegraphics[scale=0.35]{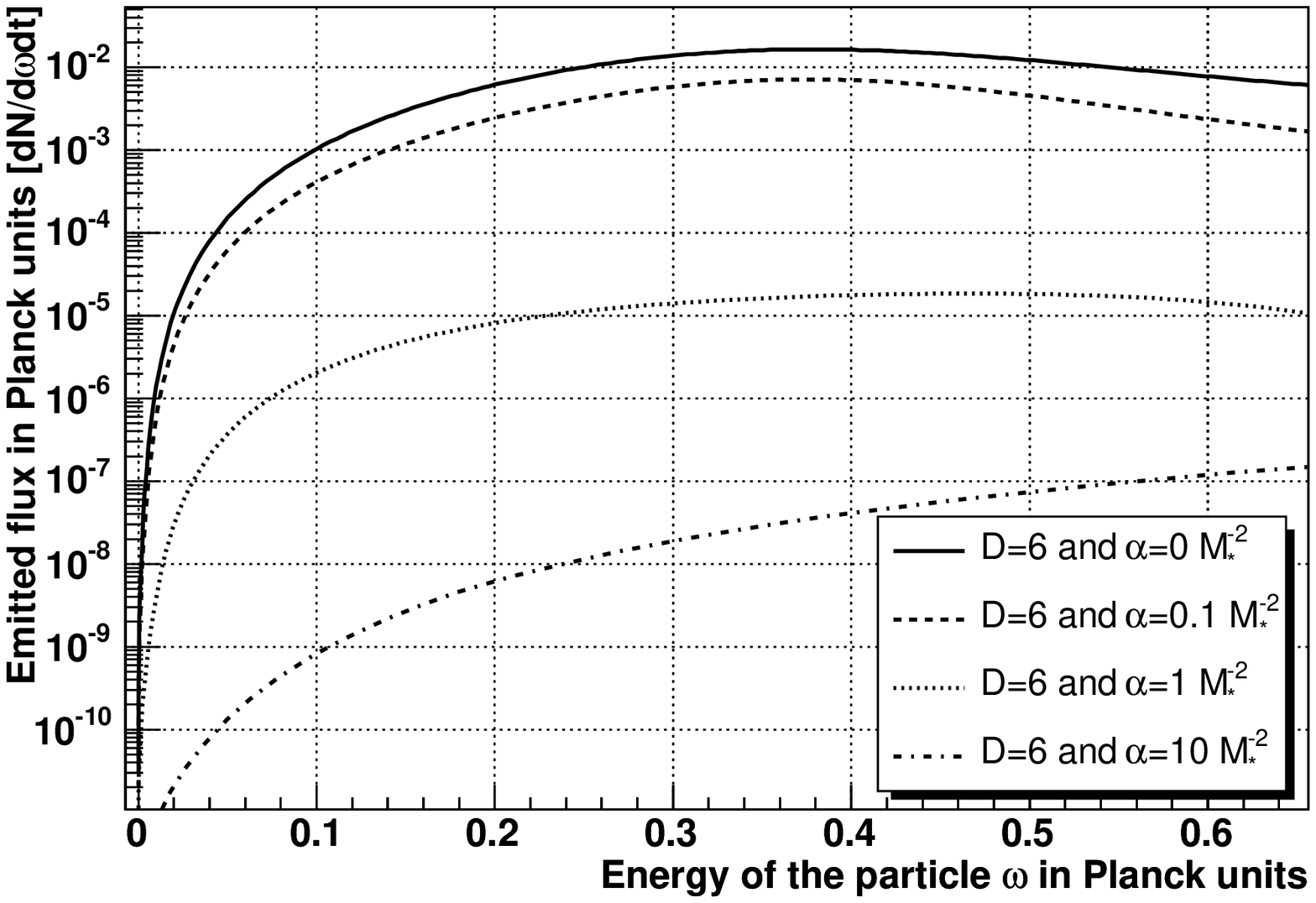}&
\hspace*{-0.5cm} \includegraphics[scale=0.35]{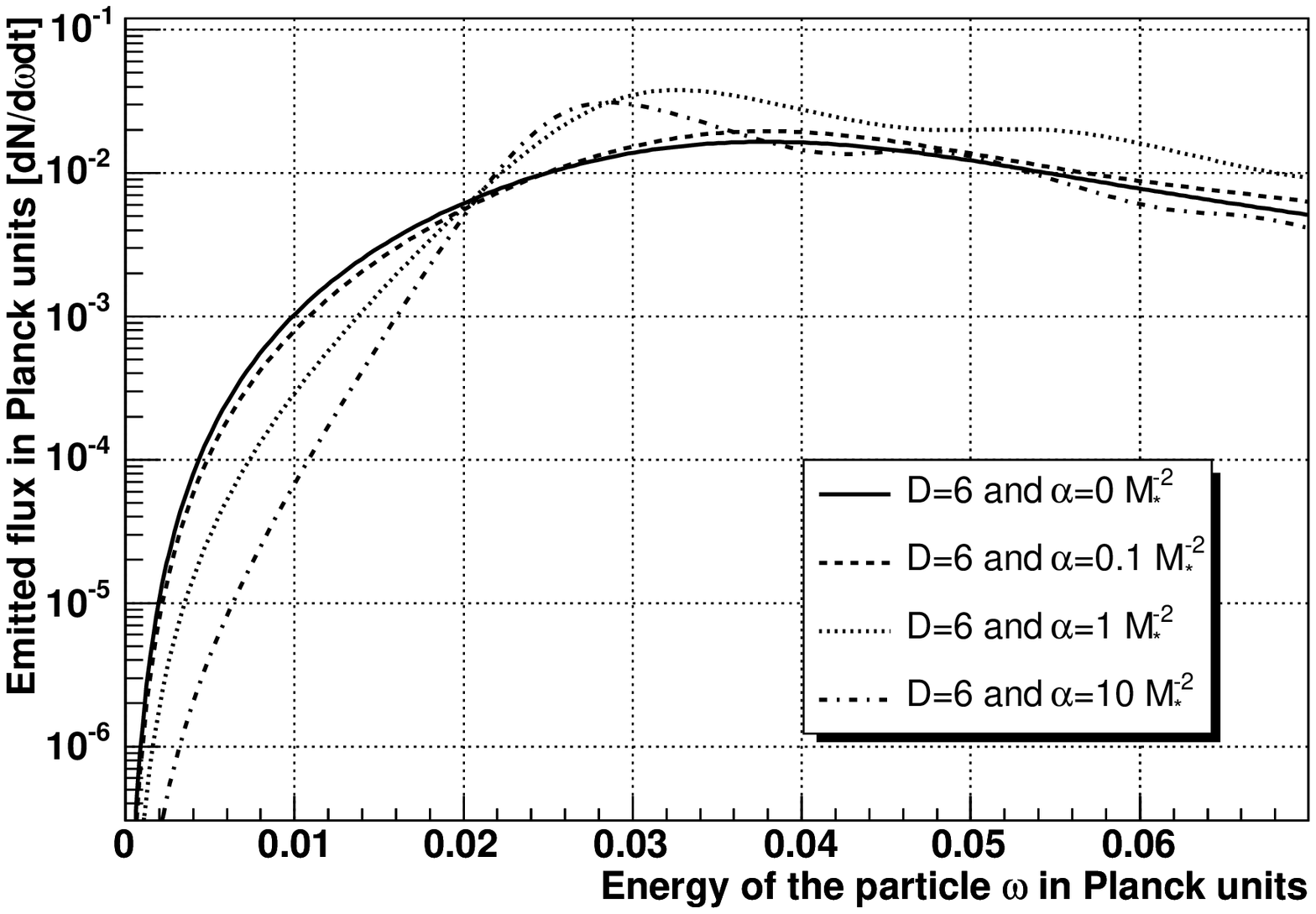}
		\end{tabular}
		\caption{Flux spectrum of gauge bosons emitted on the brane
		 by a  Schwarzschild-Gauss-Bonnet black hole as a function of 
		 the energy $\omega$. The Gauss-Bonnet coupling
		 constant takes the values $\{0,0.1,1,10\}\,M^{-2}_{*}$ and the 
		 number of dimensions is fixed at $6$. The mass of the black hole is 
                 $M_{BH}=10\,M_{*}$ on the left panel, and $M_{BH}=10000\,M_{*}$
                 on the right panel.}
		\label{bose-flu-brane}
	\end{center}
\end{figure}

For massless particles emitted on the brane, the flux spectrum, {\it{i.e.}} the number of
particles emitted per unit time and unit frequency by the black hole, is given by
Eq. (\ref{rad-spec}) evaluated for $D=4$\,:
\begin{equation}
	\frac{d^2N_s}{dtd\omega}=\frac{1}{(2\pi)^3}\frac{4\pi\sigma_{s}
        \left(\omega\right)\omega^2}{e^{\omega/T_H}-(-1)^{2s}}\,,
\end{equation}
where $s$ is the spin of the emitted field. Figures \ref{scal-flu-brane}, 
\ref{ferm-flu-brane} and \ref{bose-flu-brane} depict flux spectra for scalars,
fermions and gauge bosons, respectively, emitted by a higher-dimensional
Schwarzschild-Gauss-Bonnet black hole. As the dependence of these spectra on
the number of dimensions is similar to the one obtained in the case of a 
higher-dimensional Schwarzschild black hole, that has been exhaustively studied
in the literature already \cite{kanti2,kanti3, harris}, here we choose to focus
on the effect of the Gauss-Bonnet coupling constant instead. We thus fix the
number of dimensions at $D=6$, and assign to $\alpha$ the values
$\{0,0.1,1,10\}\,M^{-2}_{*}$. In the case of scalar fields, shown in Fig. 6,
the particle emission spectrum was also derived for four different values of the
black hole mass, {\it{i.e.}} for $\{10,100,1000,10000\}\,M_{*}$. The spectra for fermions
and gauge bosons are shown only for the two extreme mass values as their behavior
is qualitatively identical to the one seen in the scalar case.

From these plots,  we may clearly see the non-monotonic effect that the 
Gauss-Bonnet coupling constant has on the number of emitted particles. We
can also clearly see a strong dependence of the spectrum on the mass of the
evaporating black hole. A number of competing factors contribute to the
non-monotonic behaviour in terms of $\alpha$ displayed in Fig. \ref{scal-flu-brane}:
although the greybody factor for scalar fields is greatly enhanced in units of
the horizon area, the horizon radius itself shrinks under an increase in the
value of the Gauss-Bonnet coupling, when we keep the mass of the black hole
fixed; in addition, the temperature of the black hole as a function of $\alpha$
is also non-monotonic -- a simple numerical analysis shows that it initially
decreases with $\alpha$ but, after a turning point is reached, it starts
increasing again. All the above factors tend to compensate each other, with
the outcome of this process greatly depending on the black hole mass value: for
light black holes, the reduction of the horizon radius, as $\alpha$ increases,
is a significant one which, in conjunction with the decreasing temperature, leads to
the suppression of the number of particles emitted - only after the turning point
of the temperature is reached does the spectrum show some enhancement. For heavier
black holes, the effect on the value of the horizon radius, as well as the
decline of the temperature with $\alpha$, tends to become less and less important;
the increase of the absorption cross-section then prevails over all other
factors, thus leading to an enhancement of the radiation spectrum as the
value of $\alpha$ increases. 

Similar arguments hold for the case of the emission on the brane of fermions
and gauge bosons. In both cases, for low-mass black holes, the decrease of the
horizon radius and of the temperature leads to a suppression of the number
of particles emitted, even in the intermediate and high-energy regime where
the greybody factor was greatly enhanced as a function of $\alpha$; only for
large values of the Gauss-Bonnet coupling constant does the spectrum shows signs
of enhancement as a reflection again of the change in the temperature behavior.
For black holes with larger masses, the enhancement of the greybody factor
with $\alpha$, found for the intermediate and high-energy regime, prevails again
and eventually leads to a small enhancement in the number of particles emitted.
This enhancement is slightly smaller than the one found in the case of the scalar
fields as a result of the radically different behaviour of the greybody 
factor for different fields at the low-energy regime: while the enhancement,
in terms of $\alpha$, seen in this energy regime gives a boost in the
spectrum of scalar fields, its suppression for fermion and gauge bosons
restricts the enhancement of the flux spectrum at higher energy regimes.

Let us finally note here that the non-monotonic behavior of the temperature
of a Schwarzschild-Gauss-Bonnet black hole as a function of the Gauss-Bonnet
coupling constant is clearly reflected on the value of the energy peaks in the
particle spectra. For instance, the temperature of a $6-$dimensional black
hole with a mass of $10~M_{*}$, is roughly equal to $T_H=\{0.14,0.11,0.06,0.5\}\,M_{*}$
for $\alpha=\{0,0.1,1,10\}\,M^{-2}_{*}$. Focusing, for example, on the scalar
particles spectrum emitted by a black hole with a mass of $10~M_{*}$ (upper
left panel of Fig. \ref{scal-flu-brane}), we see that the peaks of the emission
curves shift towards lower or higher energy values accordingly. Moreover, they
correspond to energies lying in the range $\sim1.4\times{T}_H$ to $\sim2\times{T}_H$,
depending on the exact value of $\alpha$, which is in good agreement with the
prediction for the location of the energy peak for a pure black-body spectrum
at $\sim1.6\times{T}_H$. 

\subsection{Radiation spectrum in the bulk}
	
For massless particles emitted in the bulk, the Hawking radiation spectrum is given by\,:
\begin{equation}
	\frac{d^2N}{dt\,d\omega}=\frac{1}{(2\pi)^{D-1}}\frac{\Omega_{D-2}\,
        \sigma(\omega)\,\omega^{D-2}}{e^{\omega/T_H}-1}\,,
\end{equation}
where the spin index is omitted as only scalar fields can propagate into the
bulk (gravitons are not considered in this work). By using our numerical results 
for the bulk absorption cross-section, derived in section 4.2, we may now proceed
to derive the particle emission spectrum for emission in the bulk of scalar fields.
The absorption cross-section will be expressed in units of the $D-$dimensional
black hole area $A_{D-2}=\Omega_{D-2}r^{D-2}_H$, but, as in the previous subsection,
we choose the mass of the black hole as its characteristic parameter that remains fixed.

\begin{figure}
	\begin{center}
		\begin{tabular}{cc}
			\includegraphics[scale=0.35]{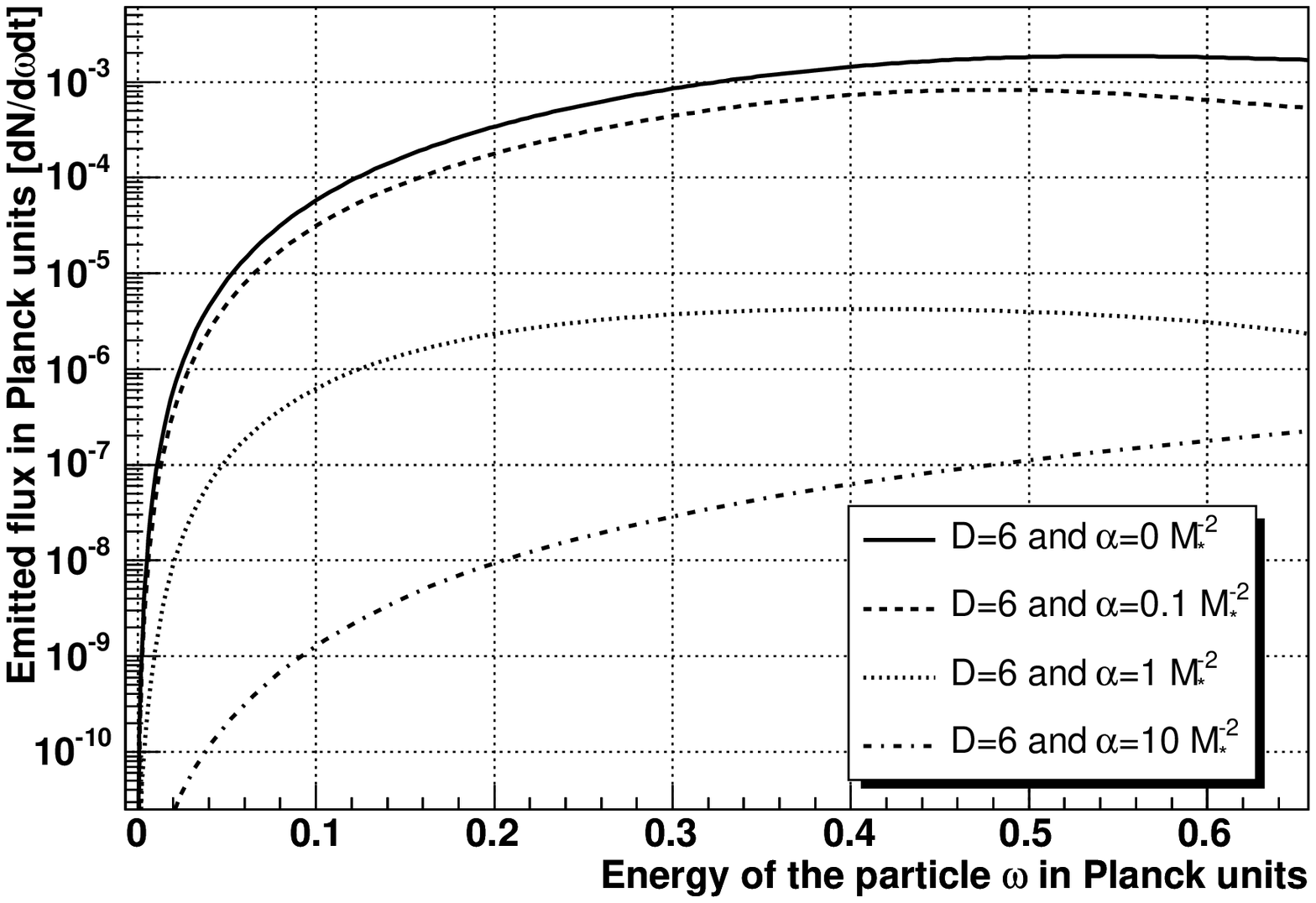}&
\hspace*{-0.5cm} \includegraphics[scale=0.35]{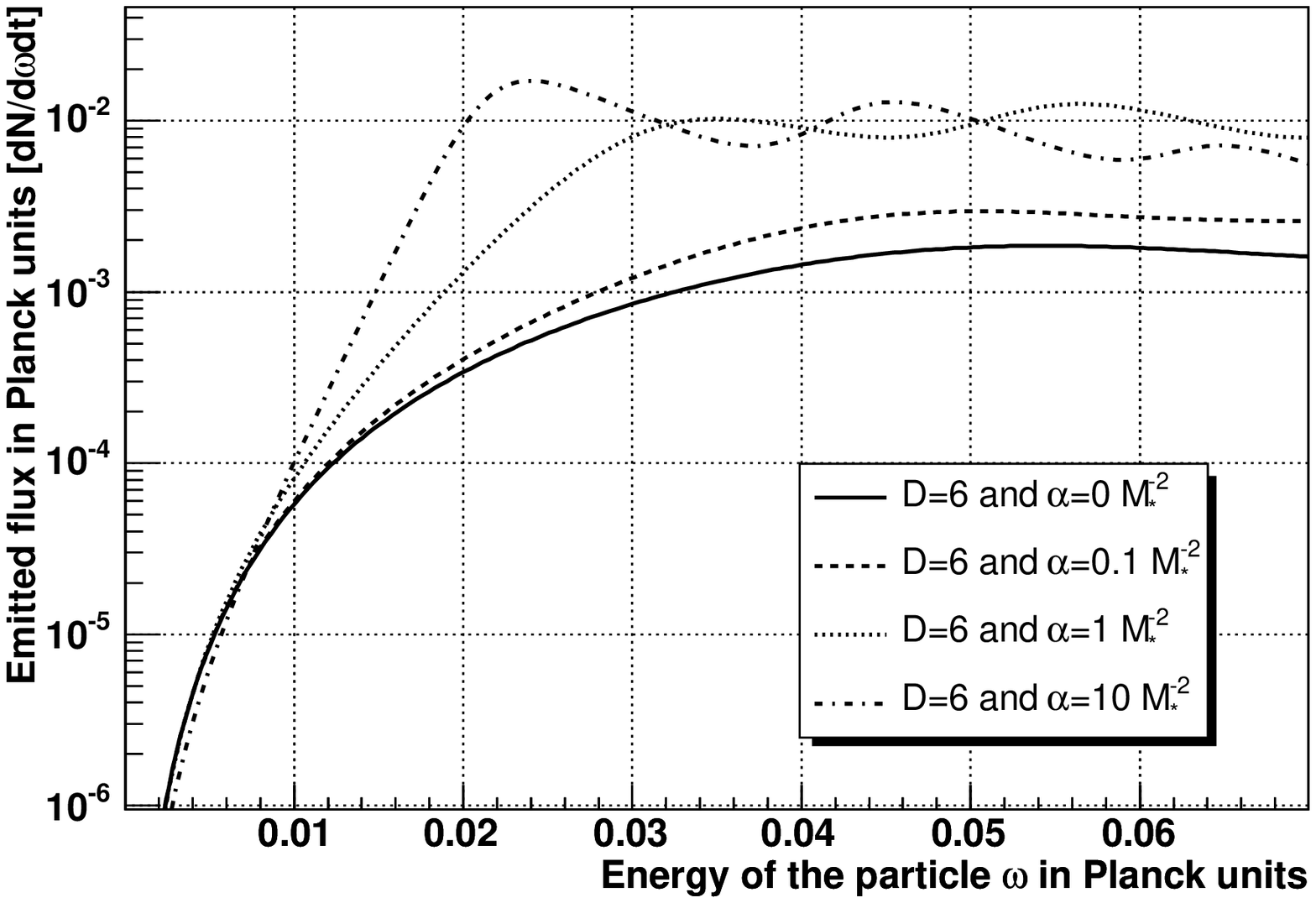}
		\end{tabular}
		\caption{Flux spectrum for scalar fields emitted in the 
		bulk by a Schwarzschild-Gauss-Bonnet black hole as a function 
		of the energy $\omega$. The Gauss-Bonnet 
		coupling constant takes the values $\{0,0.1,1,10\}\,M^{-2}_{*}$ and the 
		number of dimensions is fixed at $6$. The mass of the black hole is 
                 $M_{BH}=10\,M_{*}$ on the left panel, and $M_{BH}=10000\,M_{*}$
                 on the right panel.}
		\label{scal-flu-bulk}
	\end{center}
\end{figure}

The flux spectra for scalar particles emitted in the bulk by a $6-$dimensional black
hole, with a mass of $10~M_{*}$ and $10000~M_{*}$, are shown in the two panels of 
Fig. \ref{scal-flu-bulk}, for a set of values of the Gauss-Bonnet coupling constant
given by $\{0,0.1,1,10\}\,M^{-2}_{*}$. The spectra exhibit the same characteristics,
in terms of their behavior as functions of the Gauss-Bonnet coupling constant, as
in the case of scalar fields emitted on the brane. When the mass of the black hole
is only a few times above the Planck scale, the decrease in the horizon value and the temperature of the black hole, when $\alpha$ increases, eventually leads to the
suppression of the flux spectrum, despite the enhancement of the absorption
cross-section over the whole energy regime in units of the black hole's area.
For significantly heavier black holes though, the magnitude of this enhancement
overcomes the decrease in the area and temperature of the black hole, leading
to a significant increase in the number of particles emitted, especially for 
high values of $\alpha$. 

\subsection{Bulk-to-brane ratio}

The bulk-to-brane relative emission is of great importance for experimental 
investigations: particles emitted in the bulk cannot be detected, and the
corresponding energy will be permanently lost. It was shown in
Refs. \cite{harris,kgb} that this ratio remains lower than one for 
$D-$dimensional Schwarzschild and $D-$dimensional Schwarzschild-de Sitter black 
holes. Such results clearly favour the scenario of the experimental detection
of small black holes through the emission of Hawking radiation since most of their
energy will be spent on the emission of brane-localised particles.
However, as it was mentioned in Section 4.2, for Schwarzschild-Gauss-Bonnet
black holes, the effect of the Gauss-Bonnet coupling constant on the absorption
cross-section is more important for particles emitted in the bulk than for
particles emitted on the brane. Therefore, we can not exclude the possibility
that such black holes can decay mainly in the bulk channels instead of the brane 
channels. We expect this to happen especially in the high-energy regime, where
the cross-section is more significantly enhanced, and for heavy black holes, for
which the effect of the greybody factor dominates the spectrum. 

\begin{figure}
	\begin{center}
		\begin{tabular}{cc}
			\includegraphics[scale=0.35]{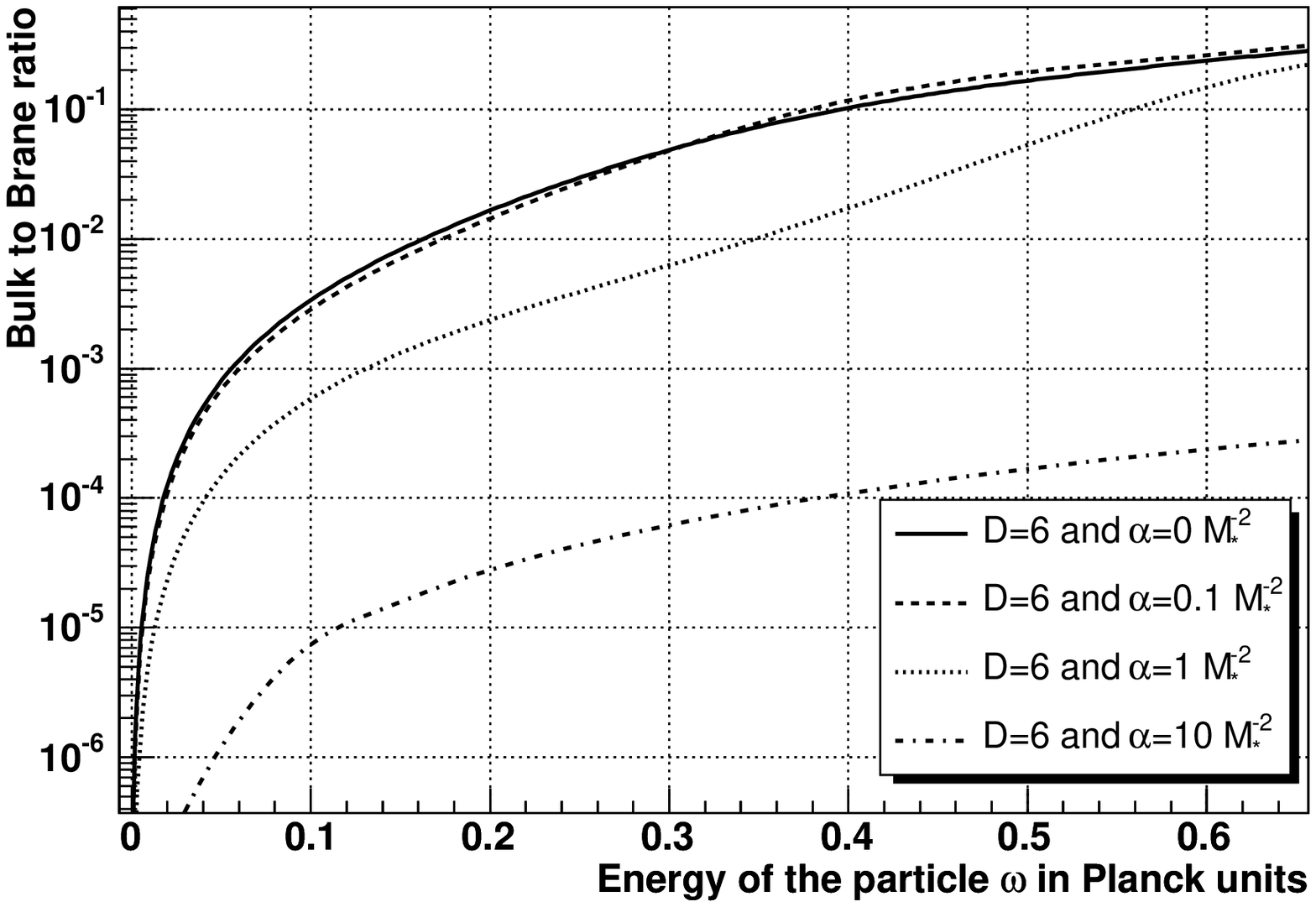}&
\hspace*{-0.5cm} \includegraphics[scale=0.35]{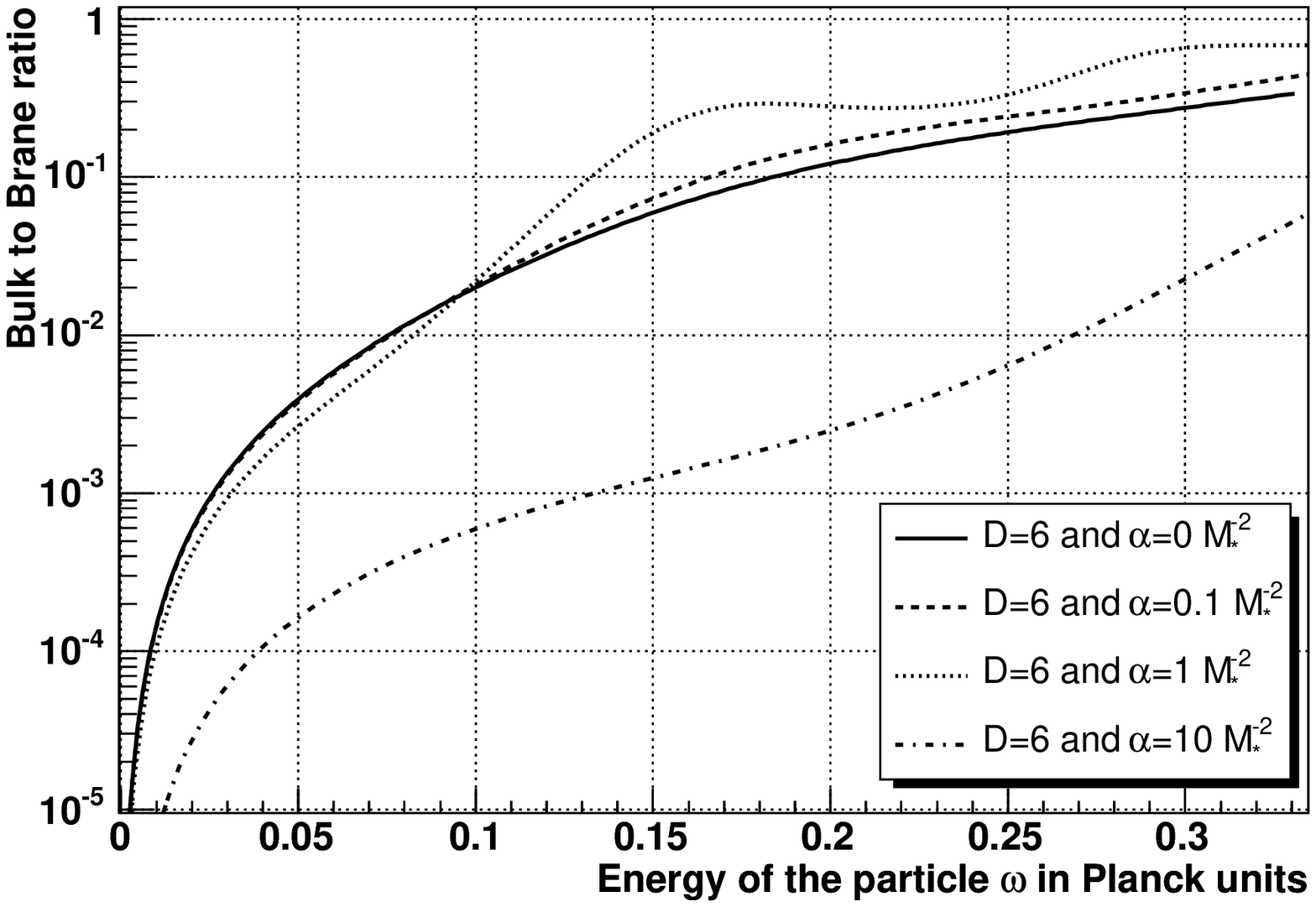} \\

			\includegraphics[scale=0.35]{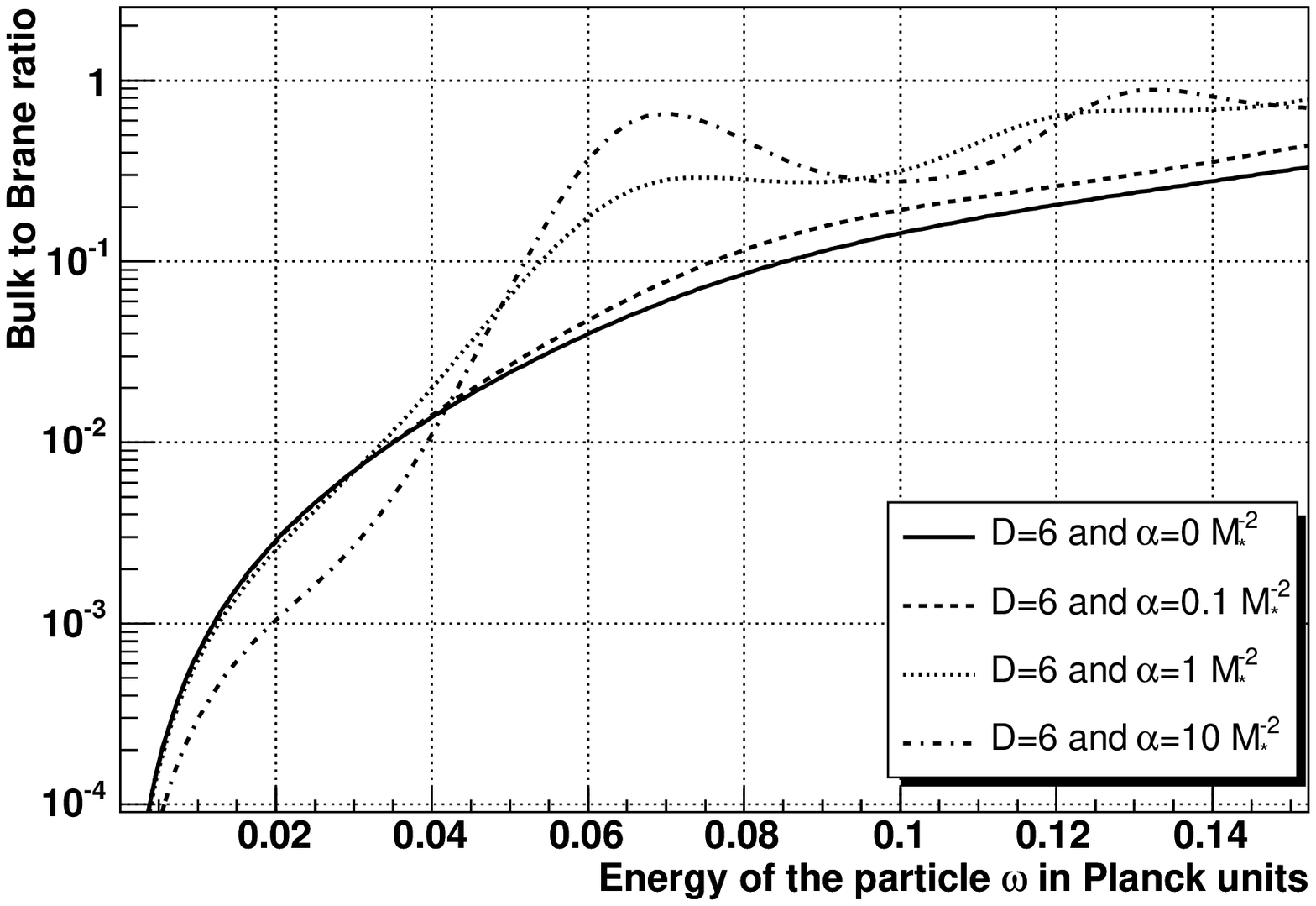}&
\hspace*{-0.5cm} \includegraphics[scale=0.35]{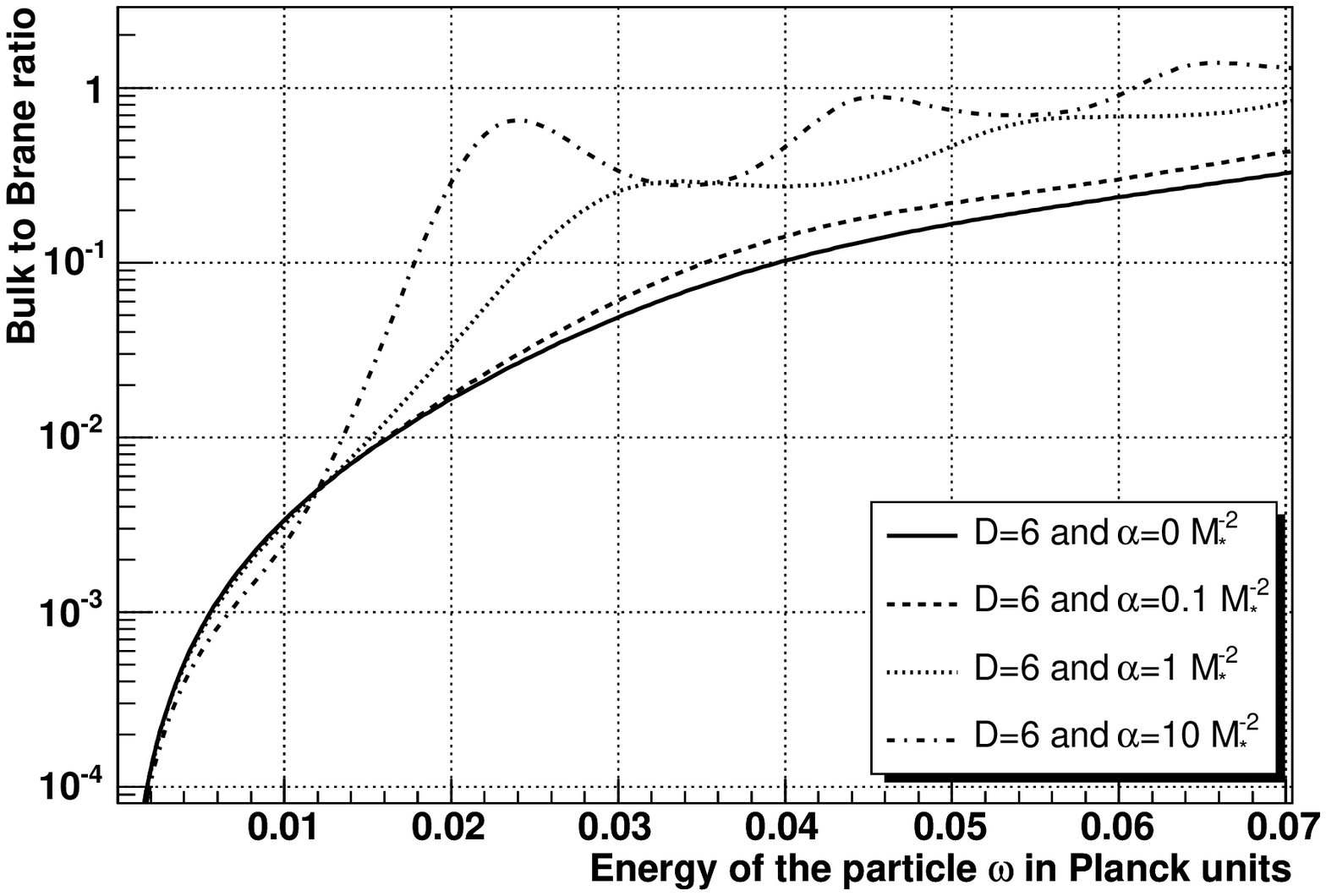}
		\end{tabular}
		\caption{Bulk-to-brane relative emission rate as a function 
		of the energy $\omega$. The Gauss-Bonnet 
		coupling constant takes the values $\{0,0.1,1,10\}\,M^{-2}_{*}$ and the
		 number of dimensions is fixed at $6$. The mass of the black hole is taken 
		to be $M_{BH}=10\,M_{*}$ (upper left), $M_{BH}=100\,M_{*}$ (upper right), 
		$M_{BH}=1000\,M_{*}$ (lower left) and $M_{BH}=10000\,M_{*}$ (lower right).}
		\label{ratio}
	\end{center}
\end{figure}

In Fig. \ref{ratio}, we present the outcome of our attempt to check this
possibility numerically by using our results for the bulk and brane particle
emission rates for scalar fields. The four panels of this figure show the
bulk-to-brane ratio for a $6-$dimensional black hole with a mass equal to
$M_{BH}=\{10,100,1000,10000\}\,M_{*}$, respectively, and for different values
of the Gauss-Bonnet coupling constant. As we see, in almost all cases, the
relative emission rate remains lower than one, although it asymptotically
approaches this value in the high-energy regime. As expected, this ratio
exceeds unity only in the case of a black hole with mass of $10000\,M_*$
and only in the high-energy part of the spectrum. Therefore, it seems highly
possible that, as we increase further the value of the black hole mass,
and for sufficiently high Gauss-Bonnet coupling constant, the bulk channel
becomes increasingly more dominant, leading eventually to a significant loss
of the black hole energy into the bulk.

\begin{figure}
	\begin{center}
		\begin{tabular}{cc}
			\includegraphics[scale=0.35]{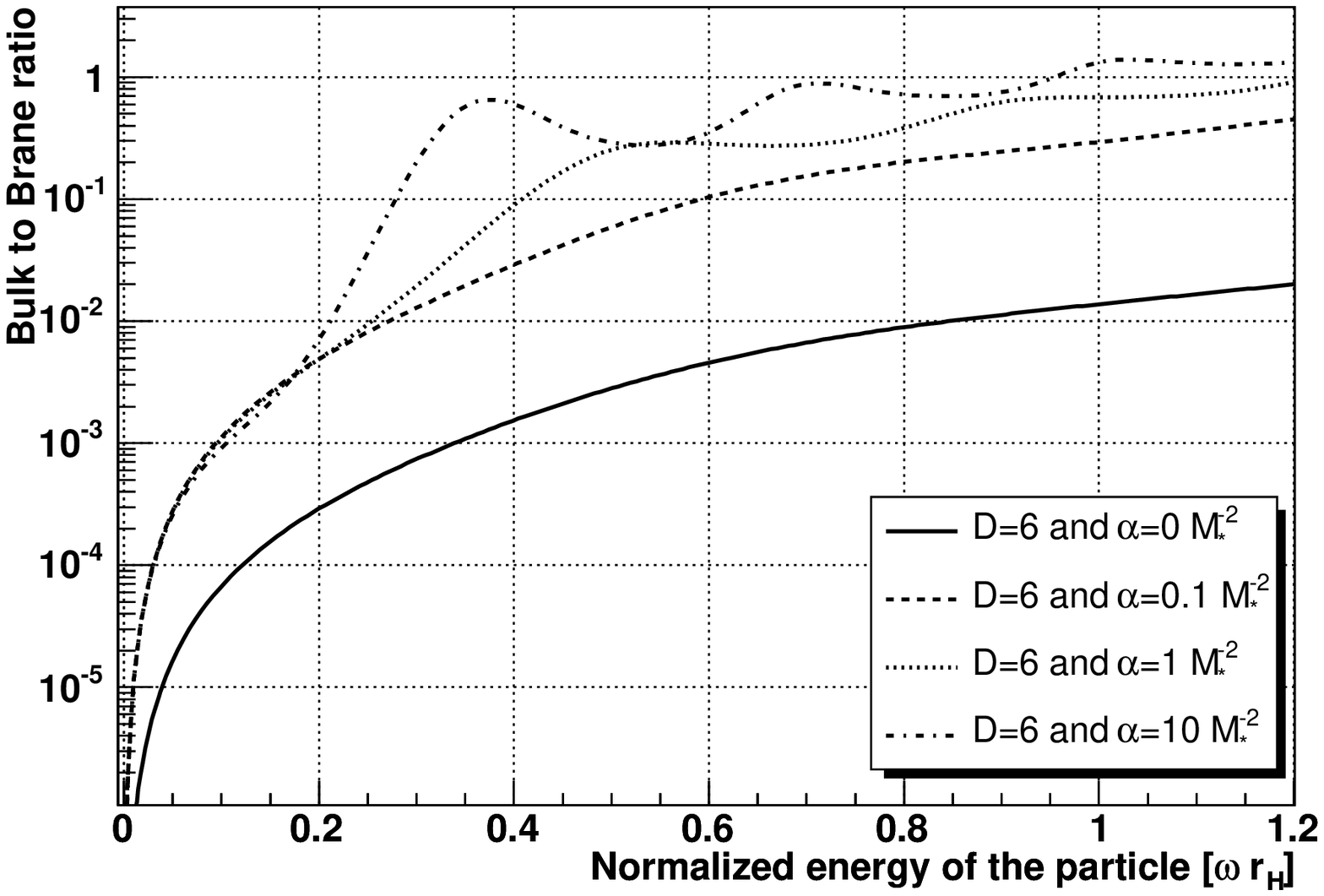}&
\hspace*{-0.5cm} \includegraphics[scale=0.35]{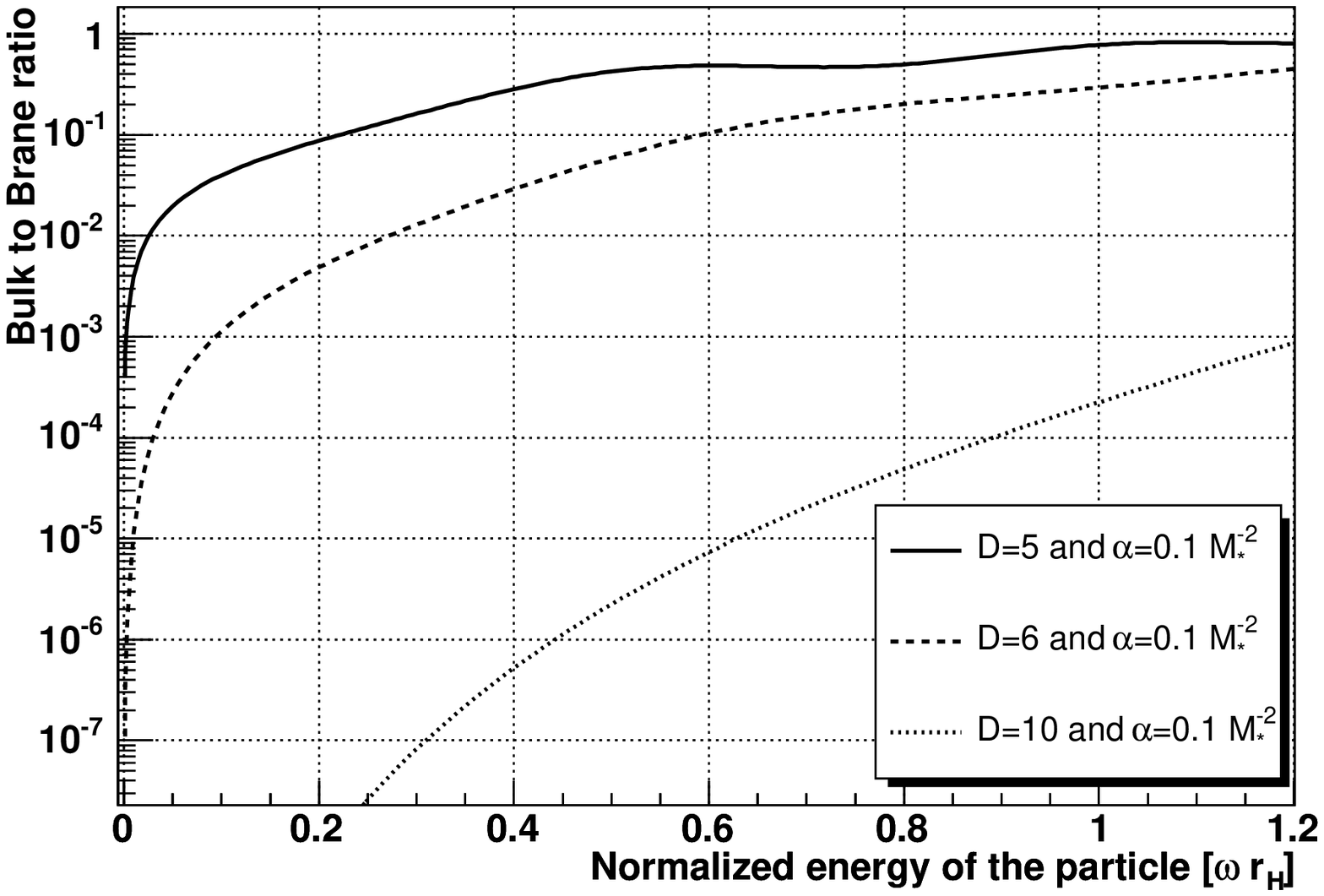}
		\end{tabular}
		\caption{Bulk-to-brane relative emission rate as a function of 
		the unidimensional parameter $\omega{r}_H$ for $6-$dimensional 
		black holes with a given size. {\it{Left Panel~:}} the 
		Gauss-Bonnet coupling constant is taken at 
		$\{0,0.1,1,10\}~M^{-2}_{*}$ and the number of dimensions is 
		fixed at $D=6$. {\it{Right Panel~:}} the Gauss-Bonnet coupling 
		constant is fixed at $0.1~M^{-2}_{*}$ and the number of 
		dimensions is taken at $\{5,6,10\}$.}
		\label{ratio-rad}
	\end{center}
\end{figure}

An upper limit on the bulk-to-brane relative emission rate can be obtained
by looking at the spectrum of black holes with fixed horizon radius instead
of a fixed mass. Figure \ref{ratio-rad}(left panel) shows the spectrum of such
a black hole.
Since $r_H$ is now kept fixed, the different curves shown in this figure for
different values of the Gauss-Bonnet coupling constant correspond to black
holes with different masses, according to Eq. (\ref{Mass}). However, there is
one case where all these curves would describe a black hole with the same mass,
and that is the case of a black hole with an infinite mass -- in that case,
any change in $\alpha$ would produce a negligible change in the black hole
mass. Therefore, Fig. \ref{ratio-rad} can be used to derive an upper limit
for the bulk-to-brane ratio. As we may see, for high values of the Gauss-Bonnet
coupling constant, the relative emission rate becomes indeed greater than one.
This value is attained over the whole high-energy regime, and can therefore
lead to a major loss of the black hole energy into the bulk (or, even the
decay of the black hole mainly through the bulk channel) via the emission of
particles with $\omega > 1/r_H$. We should, however, note here that such a 
scenario is not experimentally favored as it requires very high values for the 
Gauss-Bonnet coupling constant and black hole masses above the values that
could be reached by the next-generation colliders. The behavior of the
bulk-to-brane relative emission rate for an intermediate value of the
Gauss-Bonnet coupling constant, {\it{i.e.}} $\alpha=0.1~M^{-2}_{*}$, is shown in
Fig. \ref{ratio-rad}(right panel) as a function of the number of dimensions\,: as in
the case of a pure $D-$dimensional Schwarzschild black hole \cite{harris}
it remains lower than one and decreases for higher dimensional black holes.

\section{Conclusions}

Looking for gravitational effects beyond general relativity is a major
experimental challenge. Lovelock gravity (especially its curvature-squared 
Gauss-Bonnet term) is an interesting way of extending the Einstein-Hilbert action in
a string-inspired approach. As was shown in  \cite{bga}, Schwarzschild-Gauss-Bonnet
black holes, that could also be formed at colliders, would allow an
accurate investigation of the associated coupling constant. This paper goes beyond
the approximation made in \cite{bga} for the expression of the Hawking radiation
emission rate of these black holes, and presents the complete, exact particle
spectrum by including the missing material, {\it{i.e.}} the greybody factors.  
	
More analytically, in this paper, we have investigated the decay of a $D-$dimen\-sional
Schwarzschild-Gauss-Bonnet black hole via the emission of Hawking radiation. For this
purpose, we have computed the exact expressions for the greybody factors for scalar,
fermion and gauge boson fields emitted on the brane as well as for scalar fields
emitted in the bulk by using numerical analysis. These results have revealed that
the absorption cross-sections depend strongly not only on the energy and spin of the
emitted particle and the dimensionality of spacetime but also on the value of
the Gauss-Bonnet coupling constant. Interestingly enough, it turns out that the
effect of this coupling constant on the absorption cross-section is exactly the
opposite of the one of the dimensionality of spacetime\,: this quantity is
enhanced over the whole energy regime for scalar fields (emitted both on the brane
and in the bulk) while, for fermions and gauge fields, it is enhanced in the
intermediate- and high-energy regime and suppressed in the low-energy regime. 

With these results at hand, we proceeded to evaluate the particle emission spectra
for all types of particles. At this point, we switched from the convenient assumption
of a black hole with a fixed horizon radius to the phenomenologically realistic
one of a black hole with a fixed mass -- that allowed us to reveal and emphasize
the importance of the mass of the black hole on the behavior of the radiation spectra 
with respect to the Gauss-Bonnet coupling constant. Our results have shown that, for
light black holes, an increase in the value of the Gauss-Bonnet coupling constant
leads to the suppression of the number of particles emitted by the black hole per
unit time and unit frequency, for all types of particles. This is the result of the
ensuing decrease of the black hole's area and temperature dominating over any
enhancement exhibited by the greybody factor. However, for heavy black holes, 
it is the enhancement of the greybody factor that overcomes the small, in this
case, decrease of the black hole's area and temperature, thus leading to a spectrum
that is also predominantly enhanced with the Gauss-Bonnet coupling constant.

The bulk-to-brane relative emission rates have also been computed for the emission of
scalar fields that are allowed to propagate both on and off the brane. They were
shown to be also black-hole-mass and coupling-constant-dependent.
Whereas, for small or zero Gauss-Bonnet coupling constant, this ratio remains always
smaller than unity -- thus favoring the detection of mini black holes through their
decay mainly on the brane where our detectors are located -- this situation may 
easily change for heavy black holes and sufficiently high Gauss-Bonnet coupling constant.
Nevertheless, any significant effect of this type would demand black hole masses beyond
the order of 10000 $M_*$ and a Gauss-Bonnet coupling constant of order $10\,M_*^{-2}$. 
Therefore, for black holes with masses attainable at next-generation colliders, their
decay will take place mainly through brane channels, even if quadratic curvature terms
are taken into account in the theory. 

To conclude, let us note that the above results have been derived under the
assumption that the extra spacelike dimensions can be considered as non-compact.
This assumption might be violated if the black hole radius becomes comparable to
or larger than the size of the extra dimensions. Depending on the number of
dimensions, the latter varies in the range of a few fermis to a few millimeters
in order to obtain a Planck scale of order a TeV  without being in contradiction
with any experimental observations. As black holes produced at next-generation
colliders will not be heavier than $10~M_{*}$, their sizes will remain much
lower than a fermi, and our calculations can therefore be safely used to compute
the decay of higher-dimensional Schwarzschild-Gauss-Bonnet black holes.
Finally, our results apply for particles with wavelengths lower than the size
of the extra dimensions, which then defines a lower limit on the particle's energy. 
Therefore, the interpretation of our results found for the very low-energy part
of the derived spectra must be interpreted with care. This constraint however
does not diminish the importance of our results or the possibility of detecting
the presence of quadratic gravitational terms in the theory: the effect of
the Gauss-Bonnet term becomes manifest in a clear way over the whole radiation
spectrum, especially in the intermediate and high-energy regime.

{\bf Acknowledgments}.
The work of P.K. was funded by the UK PPARC Research Grant PPA/A/S/2002/00350.

\end{document}